\documentclass[12pt]{article}
\usepackage{amssymb,amsmath}
\usepackage[pdftex]{graphicx}

\makeatletter

\@addtoreset{equation}{section}
\def\section{\@startsection {section}{1}{\z@}{-2.5ex plus -1ex minus
 -.2ex}{1.3ex plus .2ex}{\large\bf}}
\def\subsection{\@startsection{subsection}{2}{\z@}{-2.25ex plus%
 -1ex minus -.2ex}{0.5ex plus .2ex}{\bf}}

\advance \voffset by -1.0in
\advance \hoffset by -0.7in
\def\cd{\!\cdot\!}

\newcommand {\half} [1][1] {\ensuremath{\frac{#1}{2}}}
\renewcommand{\d}{d}
\def\bx{{\mbox{\boldmath $x$}}}
\def\bz{{\mbox{\boldmath $z$}}}

\def\bu{{\mbox{\boldmath $u$}}}
\def\bv{{\mbox{\boldmath $v$}}}

\def\bpm{\begin{pmatrix}}
\def\epm{\end{pmatrix}}

\newcommand{\ZZ}{\mathbb{Z}}

\newcommand{\RR}{\mathbb{R}}
\newcommand{\CC}{\mathbb{C}}

\newcommand{\CP}{\mathbb{CP}}
\newcommand{\RP}{\mathbb{RP}}

\def\bee{\begin{equation}}
\def\eee{\end{equation}}
\def\bea{\begin{align}}
\def\eea{\end{align}}

\begin{document}
\begin{flushright}
EMPG-11-12\\
DAMTP-2011-49
\end{flushright}
\vskip 10pt
\begin{center}
{\Large \bf  Geometric Models of Matter }

\baselineskip 18pt

\vspace{1 cm}

{\bf M.~F.~Atiyah}, \\
School of Mathematics, University of Edinburgh,\\
King's Buildings, Edinburgh EH9 3JZ, UK. \\
{\tt M.Atiyah@ed.ac.uk} 
\vspace{0.4cm}

{\bf N.~S.~Manton},\\
DAMTP, Centre for Mathematical Sciences, \\ 
University of Cambridge, \\
Wilberforce Road, Cambridge CB3 0WA, UK. \\
{\tt N.S.Manton@damtp.cam.ac.uk} \\
\vspace{0.4cm}

{\bf B.~J.~Schroers},\\
Department of Mathematics, Heriot-Watt University, \\
Riccarton, Edinburgh EH14 4AS, UK. \\
{\tt bernd@ma.hw.ac.uk} \\
\vspace{0.4cm}

{ August 2011} 
\end{center}

\begin{abstract}
\noindent Inspired by soliton models, we propose a description of
static particles in terms of Riemannian 4-manifolds with self-dual
Weyl tensor. For electrically charged particles, the 4-manifolds are
non-compact and asymptotically fibred by circles over physical
3-space. This is akin to the Kaluza-Klein description of
electromagnetism, except that we exchange the roles of magnetic and
electric fields, and only assume the bundle structure asymptotically,
away from the core of the particle in question. We identify the Chern
class of the circle bundle at infinity with minus the electric charge
and the signature of the 4-manifold with the baryon
number. Electrically neutral particles are described by compact
4-manifolds. We illustrate our approach by studying the Taub-NUT
manifold as a model for the electron, the Atiyah-Hitchin manifold as a
model for the proton, $\CP^2$ with the Fubini-Study metric as a model
for the neutron, and $S^4$ with its standard metric as a model for 
the neutrino.

\end{abstract}

\baselineskip 16pt
\parskip 8 pt
\parindent 10pt

\section{Introduction}

Geometry and the quantum mechanics of particles have an uneasy
relationship which is why general relativity is hard to incorporate
into quantum field theory. String theory is an ambitious and
remarkable attempt at unification with many successes but the final
theory has proved mysterious and elusive.

Einstein and Bohr fought a long battle on this front and Bohr was
generally deemed to have won, with the Copenhagen interpretation of
quantum mechanics accepted. But Einstein's belief in the role of
geometry made a partial come-back with the adoption of gauge theories
as models of particle physics.

A more modest and limited role for geometry in nuclear physics was
proposed by Skyrme \cite{Skyrme} with the solitonic model of baryons,
i.e. proton, neutron and nuclei, now known as Skyrmions. These have 
been shown to be approximate models of the physical baryons occurring 
in gauge theories of quarks and gluons, and have been extensively 
studied \cite{BrownRho} with considerable success.

In this paper we explore a geometric model of particles which is inspired by
Skyrme's idea but with potential applications to both baryonic and
leptonic particle physics. Our model differs from the Skyrme model in
that it uses Riemannian geometry rather than field theory, and so it
is  closer in spirit to Einstein's ideas. Another key difference
is that we absorb the Kaluza-Klein idea of an extra circle dimension
to incorporate electromagnetism. However, we  exchange the roles of
electricity and magnetism relative to the standard Kaluza-Klein
approach, so that the extra circle dimension is magnetic rather than
electric, and it is the electric charge that is topologically quantised by the
famous Dirac argument. We will also, initially, ignore time and
dynamics, focusing on purely static models.

Our geometric models will therefore be 4-dimensional Riemannian
manifolds. We require the manifolds to be oriented and complete but not
usually compact: we use non-compact manifolds to model electrically
charged particles, and compact manifolds for neutral particles. Both
baryon number and electric charge will be encoded in the topology,
with baryon number (at least provisionally) identified with the
signature of the 4-manifold\footnote{The definition of signature for non-compact manifolds is reviewed  in Sect.~\ref{sigdef}.} and electric charge with minus the Chern
class of an asymptotic fibration by circles. 
In particular, the 
number of protons and the number of neutrons will therefore be
determined topologically. 

The manifolds will also have an `asymptotic' structure which
captures their relation to physical 3-space. The non-compact
manifolds we consider  have an asymptotic region which is fibred over
physical 3-space, so no additional structure is required. For the
compact (neutral) models, however, we fix a distinguished embedded
surface $X$ where the 4-manifold $M$ intersects physical 3-space. 
We also call $M \setminus X$ the inside of $M$. 
 
For single particles, the symmetry group  of rotations should fix all
of the above data. It should act isometrically on the 4-manifold and  
preserve the asymptotic structure. In the non-compact cases this means 
that it should be a bundle map in the asymptotic region, covering the
usual $SO(3)$-action on physical 3-space. In the compact cases it
should  preserve the distinguished surface $X$. In order to capture
the fermionic nature of the particles considered in this paper we also
require spin structures on the non-compact manifolds and on the
inside (in the sense defined above) of compact manifolds. Moreover,
the lift of the rotation group action to the spin bundle (over the
entire manifold in the non-compact case and over the inside in the
compact case) should necessarily be an $SU(2)$-action, and this is
what we mean by saying that our models are fermionic.

A key restriction on (the conformal classes of) our manifolds is that they are self-dual. 
Recall that the Riemann curvature is made up of the
Ricci tensor plus the Weyl tensor $W$, which is conformally invariant.
In dimension 4, $W$ is the sum of self-dual and anti-self-dual
parts
\bee
W=W^+ \oplus W^- \,.
\eee
A 4-manifold $M$ is said to be  self-dual if $W^-=0$. These manifolds are precisely those which have twistor spaces in
the sense of Penrose \cite{Penrose}. The latter are 3-dimensional
complex manifolds $Z$ with a real $S^2$-fibration over $M$. The complex
structure of $Z$ (together with a real involution which is the
antipodal map on each $S^2$) encodes the entire conformal structure
of $M$. Even the Einstein metric can be captured by complex data on $Z$.
 
A simply-connected self-dual 4-manifold, for which the Ricci tensor 
is also zero, is a hyperk\"ahler manifold, whose structure group 
reduces to $SU(2)$. It is a complex K\"ahler manifold for an 
$S^2$-family of complex structures, but for any of these complex 
structures, and for the complex orientation, it would be anti-self-dual. 
Since we want self-dual manifolds we choose the opposite orientation.

While some of our particles, including the proton, will be modelled by
hyperk\"ahler manifolds, we do not want to be so restrictive. Instead
we will only require our 4-manifold models of particles to be
self-dual and Einstein, so there can be a non-zero scalar curvature.
Our model for the neutron is of this type, distinguishing it from the
proton. The neutron will of course also have electric charge zero.

We should point out that reversing the orientation of a 
4-manifold turns a self-dual manifold into an anti-self-dual one.
This should be interpreted as giving the geometric model of an
anti-particle. The existence of anti-particles follows from $CPT$ 
invariance, and our models are compatible with this.

Self-dual 4-manifolds are, in many ways, the 4-dimensional analogue of
Riemann surfaces, with $H_2$ replacing $H_1$ in homology. In
particular there are theorems \cite{DonaldsonFriedman} which assert 
that such manifolds admit connected sums  although,  unlike in the case 
of Riemann surfaces, there are restrictions on when this is possible.
Such connected sums model composite objects, like nuclei.
Although we focus at present on static particles we do envisage a
deformation theory, using the moduli space of self-dual manifolds,
which could underlie particle interactions.

Fortunately a lot is now known about self-dual 4-manifolds with many
metrics explicitly calculated. This makes it possible to put forward
some definite models for the proton and neutron.
Even though our ideas are inspired by Skyrme's theory of baryons, it
turns out that geometric models of leptons, i.e. the electron and
(electron-)neutrino, are even simpler, and we shall describe them too 
in the class of self-dual manifolds. Thus, somewhat surprisingly, our 
framework of self-dual manifolds allows us to describe baryons and 
leptons in a unified fashion.

The language and spirit of our model for particles is close to that 
of general relativity and suggests the possibility of a unification 
with gravity, but we do not address this issue here. In particular, we do
not specify an action functional. Instead, we focus on how our model
describes general features of particles such as their various
quantum numbers. We are aware that a description of
elementary particles as 4-dimensional Riemannian manifolds is
radically different from established treatments in terms of quantum
field theory. What we aim to show in this paper is that such a
geometric approach is possible, and that it has some surprising and
attractive features, such as the possibility of describing the
electron and the proton in one framework. While we do propose
definite identifications of certain 4-manifolds with specific
particles in Sects.~3-5 of this paper, these should be seen as
illustrations of the geometric approach, not necessarily as final
proposals.
 
The paper is organised as follows. In Sect.~2 we outline the
genesis of our geometric models of particles, starting with the Skyrme
model of baryons. Electrically charged particles are necessarily
described by non-compact 4-manifolds in our approach, and
in Sect.~3  we explain how to model the electron and proton in
terms of the Taub-NUT and Atiyah-Hitchin manifolds, respectively. 
Neutral particles are described by compact 4-manifolds, and this is 
discussed in Sects.~4-5. We propose $\CP^2$ as a model for the neutron and
$S^4$ as a model for the neutrino. These are the simplest choices, but 
we also discuss  some more sophisticated versions. In Sect.~6 we
describe how our particle models glue into empty space, and how the 
particles may interact with each other. Sect.~7 contains an outline 
of how our geometric models capture
the spinorial nature of the particles they  describe. In Sect.~8 we give the
dictionary which translates topological properties of 4-manifolds
into the electric charge and baryon number of particles, and discuss in
some detail how these charges are related to fields and densities
used in conventional Lagrangian models of particle physics. Sect.~9
contains our conclusion and some ideas for follow-up work. Conventions and
calculations are collected in appendices A,B and C.
 
\section{From Skyrmions to 4-manifolds}

We begin by spelling out in detail how the Skyrme model suggests 
our 4-manifold model. The Skyrme model is based on a group-valued 
field from $\RR^3$,
\bee
\label{skyrmefield}
U : \RR^3 \rightarrow G \,,
\eee
where the Lie group $G$ is usually taken to be $SU(2)$, and $U(\bx)
\rightarrow  1$ as $|\bx| \rightarrow \infty$. The degree of $U$ as
a map $S^3 \rightarrow SU(2)$ is identified with baryon number.
The minima of the Skyrme energy, for each baryon number, are called
Skyrmions.

Skyrmions are free to rotate, both in physical space and through
conjugation by elements of $SU(2)$. Quantising this motion gives the
Skyrmions spin and electric charge. The proton and neutron, for
example, are distinct quantum states of the essentially unique 
Skyrmion of degree 1.

In \cite{AtiyahManton} it was shown how to generate such Skyrme
fields naturally by starting with an $SU(2)$ Yang-Mills gauge field 
on $\RR^4$ and calculating the holonomy along the 4th direction. Suitable
asymptotic behaviour on $\RR^4$ guarantees a well-defined map $U$.
Although this construction does not preserve the respective energy
functionals it does provide a good way of using instantons on
$\RR^4$ (i.e. self-dual gauge fields) to construct approximate minima 
of the Skyrme energy. It 
also identifies instanton number with the Skyrme degree. See also 
the recent papers \cite{SakaiSugimoto,Sutcliffe} where the difference 
between the Yang-Mills and Skyrme energy functionals is interpreted 
as due to an infinite tower of mesons.

Since the Yang-Mills energy functional in dimension 4 is conformally
invariant we could replace the decomposition
\bee
\RR^4 = \RR^3 \times \RR^1 
\eee
by
\bee
S^4 \setminus S^2 = H^3 \times S^1 \,,
\eee
where $H^3$ is hyperbolic 3-space. In fact we can vary the curvature
of $H^3$ provided we rescale the circle $S^1$ the opposite way, so
that large circles correspond to almost flat $H^3$.
We can now fix a gauge field on $S^4 \setminus S^2$ and take the
holonomy round the circles. There are some technicalities (due to
base-points) which we shall ignore but basically we expect to end up
with a Skyrmion on $H^3$, an idea which has been explored in
\cite{AtiyahSutcliffe}.

Now replace $H^3 \times S^1$ by any Riemannian 4-manifold $M$ which is
asymptotically fibred by circles over $\RR^3$. This is the kind of
Kaluza-Klein 4-manifold we are going to consider. An $SU(2)$ gauge
field on $M$ would then give a Skyrme field on `the quotient of $M$ by
$S^1$'. Since we do not want to assume there is a global circle
fibration, this Skyrme field will only be defined asymptotically 
outside some `core'. But an oriented 4-manifold has two natural
$SU(2)$ bundles over it, the two spin bundles $S^+$ and $S^-$
(assuming $M$ is a spin-manifold, i.e. $w_2(M)=0$). Picking one of
these, say $S^+$, we then get from the connection on $S^+$ a 
natural construction of an asymptotic Skyrme field on $\RR^3$.

This is roughly the genesis of our idea to model particles by 4-manifolds,
but the topology of these asymptotic Skyrme fields does not quite fit, 
and would not give integer baryon numbers as defined in our model. The
reason lies in a fundamental difference between the topology of gauge
bundles which can have arbitrary instanton number (or second Chern
class) and the topology of tangent bundles, where there are
divisibility theorems. For example the first Pontrjagin class of a
4-manifold (compact and oriented) is divisible by 3 and then gives the
signature. An example is $\CP^2$ which has
signature 1 and Pontrjagin class 3 (times the generator of $H^4(\CP^2)$).

In the Skyrme model the basic idea is that baryon number is identified
with the degree of the map $U$ in \eqref{skyrmefield}, or
equivalently with the instanton number (or second Chern class) of the
$SU(2)$ bundle over $\RR^4$. This differs from the 4-manifold model we
want to explore, where baryon number is identified with the signature
of the 4-manifold. The signature is additive under taking connected
sums of 4-manifolds \cite{AtiyahSinger}, and this captures the
additivity of baryon number for composites of particles, for example,
fusion of nuclei. The integrality of
the signature is linked to it being an index of an elliptic
operator. This means we are in the realm of K-theory rather than
cohomology. A consequence of this change of viewpoint is that the
geometry of the 4-manifold model is important for us, but we will not
try to define a global 3-dimensional Skyrme field $U$.

Recall that in the Skyrme model, baryon number is cohomological and
electric charge arises at the quantum level. For our 4-manifold model,
electric charge is cohomological, arising, as already explained, from 
the first Chern class of the asymptotic $S^1$-fibration, while baryon 
number as just indicated should be seen as an index.

To sum up our discussion, we see that our model goes beyond the Skyrme
model in aiming to understand topologically both the basic integer 
physical invariants, baryon number and electric charge. The two models 
are different, but possibly dual in a suitable sense. We hope to
explore this in detail at a later stage.
 
\section{Models for the electron and proton}
\label{protonelectron}

Models for the basic particles should exhibit a high degree of
symmetry and we expect the rotation group $SO(3)$ of $\RR^3$, or
its double cover $SU(2)$, to act as isometries. For electrically
charged particles, we take our geometric models to be non-compact
hyperk\"ahler manifolds. We also assume that the volume grows with
the third power of the radius, to allow for an interpretation of the
asymptotic region in terms of physical 3-space. As recently shown in
\cite{Minerbe}, this forces the hyperk\"ahler manifold to be ALF. 
We are therefore looking for rotationally symmetric and complete 
ALF hyperk\"ahler manifolds. There are just two possibilities:
\begin{enumerate}
\item The Taub-NUT manifold \cite{Hawking,EGH} depending on a positive
parameter $m$ (interpreted as mass in the gravitational context). For 
brevity we denote it by TN.
\item The Atiyah-Hitchin manifold, the (simply-connected double cover
of the) moduli space of centred $SU(2)$-monopoles of charge two
\cite{AtiyahHitchin}. For brevity we denote it by AH.
\end{enumerate}
Note that we could also single out TN and AH among non-compact, 
complete and rotationally symmetric hyperk\"ahler manifolds by
demanding that the $SU(2)$- (or $SO(3)$-) action rotates the complex
structures, see our discussion following \eqref{gendual}. This turns
out to play a role in recovering the  usual rotation action on 
physical 3-space in the asymptotic region of  our geometric models, as  discussed   in Sect.~\ref{spinsect}. 

Both TN and AH can be parametrised in terms of a radial coordinate $r$
and angular coordinates on $SU(2)$ (for TN) or $SO(3)/\ZZ_2$ (for
AH). Details are given in  Appendix~\ref{AHTNdetails}. In terms of
the left-invariant 1-forms defined in \eqref{linv}, the metrics of both
TN and AH can be written as
\begin{align}
\label{athi}
ds^2 = f(r)^2dr^2 + a(r)^2\eta_1^2 +b(r)^2\eta_2^2 +c(r)^2\eta_3^2 \,, 
\end{align}
with the functions $f,a,b,c$ satisfying the self-duality equations
\bee
\label{dual}
\frac {2bc}{f}\frac { da}{dr} = (b-c)^2 -a^2 \,, \qquad 
\mbox{+ cyclic} \,, 
\eee
where $\mbox{+ cyclic}$ means we add the two further equations
obtained by cyclic permutation of $a,b,c$. We adopt the convention 
\bee
f(r)=-\frac  {b(r)} {r} \,, 
\eee
where (for reasons that will emerge later) the radial coordinate $r$ has the range $[0,\infty)$ for
TN and $[\pi,\infty)$ for AH. The self-duality equations \eqref{dual} become
\begin{align}
\label{abcsyst}
\frac {da}{dr} &=\frac{1}{2rc}(a^2 -(b-c)^2) \,, \nonumber \\
\frac{db}{dr}&=\frac{b}{2rca}(b^2-(c-a)^2) \,, \nonumber \\
\frac{dc}{dr}&=\frac{1}{2ra}(c^2-(a-b)^2) \,.
\end{align}

This system has  solutions in terms of elementary functions
\bee
\label{TNcoef}
a(r)=b(r) = r\sqrt{\epsilon +\frac {m}{r}} \qquad \mbox{and}\qquad
c= \frac {m}{\sqrt {\epsilon +\frac m r}} \,,
\eee 
with  parameters $\epsilon, m>0$, associated to the TN manifold. The 
topology is that of $\RR^4$, and as $\epsilon \rightarrow 0$
the metric tends to the flat metric. For $\epsilon >0$ 
the manifold is asymptotic to an $S^1$ fibre-bundle over $\RR^3$
with the length of the circle being $4\pi m/\sqrt{\epsilon}$. There is a
$U(1)$ symmetry acting along the fibres, with just one fixed point at
the origin, $r=0$. The whole isometry group is $U(2)$.
As $\epsilon \rightarrow 0$ the $U(1)$-action becomes the scalar action on
$\CC^2$. The complex orientation of $\CC^2$  determines the
orientation of TN as a self-dual manifold; this is opposite to the
orientation given by any of the complex structures in the
hyperk\"ahler family, see Appendix~\ref{signsTNAH} for a discussion.
At infinity, the $U(1)$-action gives the standard Hopf line bundle
over $\CP^1$ with Chern class $+1$; details are given in 
Appendix~\ref{conventions}. 

The TN metric with coefficient functions \eqref{TNcoef} has the
following behaviour under scaling by non-vanishing real numbers
$\alpha,\beta$:
\bee
r \rightarrow \frac{\beta}{\alpha} \, r \,, \quad m\rightarrow \alpha
\beta m \,, \quad \epsilon \rightarrow \alpha^2 \epsilon \,, \qquad
\text{then} \quad ds^2 \rightarrow \beta^2 ds^2 \,. 
\eee
We use rescaling by $\alpha$ to set $\epsilon=1$, and rescaling by
$\beta$ to set $m=2$ from now on. This amounts to picking a unit of
length for the radial coordinate $r$ and to fixing an overall scale
for the metric. Our choice is motivated by the asymptotic form of
the AH metric, to be discussed below. Note that, with  this choice,
the length of the asymptotic circle, in the length units chosen, is
$8\pi$.

The solution which gives rise to AH has the {\em asymptotic} form, for
large $r$,
\bee
\label{AHcoef}
a(r) \sim b(r) \sim r\sqrt{1-\frac {2}{r}} \qquad \mbox{and}\qquad
c(r) \sim -\frac {2}{\sqrt {1 -\frac 2 r}} \,.
\eee
These asymptotic expressions, a TN metric with $\epsilon=1, m=-2$, 
also satisfy \eqref{abcsyst}. However $a(r)$ is not actually equal to 
$b(r)$, and $r$ only extends down to $\pi$. 
For $r$ near $\pi$, \eqref{AHcoef} is a poor approximation. Instead,  
one finds the leading terms
\bee
\label{AHsmallr}
a(r) \sim 2(r-\pi) \,, \quad b(r) \sim  \pi+ \frac 1 2 (r-\pi) \,, 
\quad c(r) \sim -\pi +\frac 1 2 (r-\pi) \,,
\eee
which we will need later in this paper.

The manifold AH is the complement of $\RP^2$ (the real projective
plane) embedded in $\CP^2$, and the complex orientation of $\CP^2$ 
determines the orientation of AH as a self-dual manifold. As for TN 
this is opposite to the orientation given by any of the complex 
structures in the hyperk\"ahler family; see our discussion in
Appendix~\ref{signsTNAH}. AH has an $SO(3)$ symmetry with just one
2-dimensional orbit at $r=\pi$, which is a minimal 2-sphere. We refer
to this minimal 2-sphere, which is the totally imaginary conic in $\CP^2$ and
determined by $z_1^2+z_2^2+z^2_3=0$ in the homogeneous coordinates
introduced in Sect.~\ref{cp2sect}, as the  core.  Asymptotically,  
the manifold is fibred by circles. As further discussed below,
neither the circles nor the base space of this asymptotic fibration 
are oriented because of a $\ZZ_2$-identification, given explicitly in  
\eqref{ident}.

The manifold TN is usually interpreted as the geometry of a Dirac
monopole at the origin \cite{Pollard,GrossPerry,Sorkin}. For us, with
electric and magnetic charges reversed, it has to be interpreted as an
electrically charged particle. Since the signature of TN is zero 
(we discuss this further in Sect.~\ref{enfluxsection}) the particle 
is leptonic. We therefore interpret TN as a model for the electron. 
Down on $\RR^3$, after factoring by $U(1)$, any 2-sphere
surrounding the origin has an electric flux emerging from it due to
the electron, which carries charge $-1$. This implies that there is a
sign change in going from the Chern class to the electric charge.

The manifold AH has the opposite asymptotic behaviour with a sign
change for $m$ and an orientation change (see
Appendix~\ref{conventions}) and so would lead us naturally to expect
electric charge +1. Also, the topology at the core is different, with a
2-sphere instead of a point. As a result, AH has signature 1 (again
discussed further in Sect.~\ref{enfluxsection}) and looks like
the model we want for the proton (rather than the positron).

However things are not quite that simple, as we shall now explain.
The `asymptotic boundary' of AH  is not $S^3$ as for TN but the
boundary of a tubular neighbourhood of $\RP^2$ in $\CP^2$, which is
$S^3$ divided by a cyclic group of order 4. Moreover the base of this
unoriented circle fibration is $\RP^2$, not $S^2$, and is
non-orientable. This means that the 3-manifold which is the base of 
the asymptotic fibration is not $\RR^3$, and has a fundamental group 
of order 2. It is not orientable.

This might seem to be a disaster, but we shall argue that, while
unexpected, it is not as bad as it looks. The most convincing
argument in its favour is to show that the electric charge is
well-defined and equal to +1 as hoped. This is done in detail in
Appendix~\ref{conventions}. The lack of orientability in physical 3-space 
should be thought of as follows. The lack of orientation in
$\RR^3$ locally is compensated by a corresponding ambiguity in the
sign of electric charge (non-orientability of the circle fibres).
Physically the geometric orientation is not felt. 

\section{The neutron}
\subsection{Complex projective plane}
\label{cp2sect}

Having put forward a definite proposal for the proton we now have to
face the neutron. Since the neutron has no electric charge any
non-compact model would need to have a trivial asymptotic circle 
fibration. The 4-manifold should have signature 1 and it should resemble 
the AH model of the proton in its $SO(3)$ orbit structure. However,
the latter requirement rules out asymptotically trivial circle bundles 
over physical 3-space since the generic $SO(3)$ orbits would be 
2-dimensional in that case. We therefore consider compact 4-manifolds.  
In fact, there is an obvious choice which is just the complex 
projective plane $\CP^2$ with its Fubini-Study metric (and its natural 
complex orientation). This is a self-dual manifold of positive scalar 
curvature.

$\CP^2$ has even more symmetry than we need, since it is acted on by $SU(3)$. 
The rotation group $SO(3)$ sits inside as the subgroup that preserves
the real structure given by complex conjugation.  This preserves
$\RP^2$ as in the case of the proton, so we fix this $\RP^2$ as the 
distinguished surface where the 4-manifold intersects physical
3-space, thus breaking the symmetry to $SO(3)$. The global $SU(3)$ 
symmetry might give us a link to quarks, but this remains to be explored.

The Fubini-Study metric is often written in coordinates which
exhibit the invariance under $U(2) \subset SU(3)$. This brings out
the parallels with TN \cite{GibbonsPope} but is not the symmetry we
want in our neutron model. For the interpretation of $\CP^2$ as a
neutron and  for a comparison with the AH model of the proton, we need
to write the Fubini-Study metric in coordinates adapted to the
$SO(3)$-action, which is discussed in \cite{BouchiatGibbons} and
\cite{DancerStrachan}. We write the results of \cite{BouchiatGibbons} 
in the conventions used in our discussion of the AH metric.
 
In terms of homogeneous coordinates $ \bz \in \CC^3\setminus\{0\}$
(with the identification $\bz \simeq \lambda \bz$, $\lambda \in
\CC^*$) the Fubini-Study metric on $\CP^2$  is
\bee
ds^2 = \frac{|\bz^2| d \bz^\dagger  d\bz - \bz^\dagger d\bz
\bz^td\bar{\bz}}{|\bz|^4} \,.
\eee
For calculations we can fix $|\bz| =1$ and parametrise 
\bee
\label{ourchoice}
\bz =e^{i\alpha} R \bz_0 \,,
\eee
where $R\in SO(3)$ can, in turn, be parametrised in terms of Euler
angles as shown in Appendix B.1. The reference vector $\bz_0$ (which
depends on one parameter) should be a unit vector and we can assume,
by adjusting the phase $\alpha$ if necessary, that its real and
imaginary parts are orthogonal. For our purpose, it is convenient to
single out the 1-axis and pick
\bee
\label{referencevec}
\bz_0 = \begin{pmatrix} 0 \\  a_2 \\ i a_3 \end{pmatrix}, 
\quad  a_2^2 + a_3^2=1 \,.
\eee
Then  we find  the following expression for the Fubini-Study metric in
terms of the left-invariant forms \eqref{linv} on $SO(3)$ (for details
of an analogous calculation we refer the reader to
\cite{BouchiatGibbons}):
\bee
\label{cp2first}
ds^2 =d a_2^2 + da_3^2 + (a_3^2-a_2^2)^2 \,\eta_1^2  + a_3^2\eta_2^2
+a_2^2\eta^2_3 \,.
\eee
Parametrising
\bee
\label{anglepick}
a_2=\cos\left(\frac{\rho}{2}+\frac{\pi}{4}\right), 
\quad a_3=\sin\left(\frac{\rho}{2} + \frac{\pi}{4}\right), 
\quad \rho \in \left[0,\frac{\pi}{2}\right], 
\eee
we obtain, finally, the Fubini-Study metric in the form 
\bee
\label{cp2second}
ds^2 =\frac 1 4 d \rho^2 + \sin^2\rho \,\eta_1^2  +
\sin^2\left(\frac{\rho}{2}+\frac{\pi}{4}\right) \eta_2^2 +
\cos^2\left(\frac{\rho}{2} +\frac{\pi}{4}\right)\eta^2_3 \,.
\eee

There is a simple interpretation of the geometry of $\CP^2$ and its
orbit structure in terms of orientated ellipses up to scale
\cite{AtiyahMantonCMP}, which is useful for comparison with the AH
metric. As already exploited above, we can adjust the phase in the
homogeneous coordinate $\bz=\bu+ i\bv$ (no longer fixed to satisfy
$|\bz|=1$) so that the real vectors $\bu$ and $\bv$ are orthogonal: if
$\bz^2 =0$ this is automatic, and if $\bz^2\neq 0$ we multiply by  a
unit complex number to set Im$(\bz^2) =2\bu\cdot \bv=0$ and Re$(\bz^2)
=\bu^2 - \bv^2<0$ (we pick the negative sign to agree with the
choice made in \eqref{anglepick} above). We can interpret $\bv$ and
$\bu$ as the major- and minor-axis of an ellipse. This ellipse is only
determined up to scale (we can still rescale $\bz$ by any positive
real number) but it is orientated. The totally degenerate case
$\bz=0$ is excluded by the definition of homogeneous coordinates, but
circles ($\bz^2=0 $ or $|\bu| =|\bv|$) and lines ($|\bu|=0$)  can
occur.
 
In terms of our parametrisation \eqref{ourchoice}, the reference
vector \eqref{referencevec} and the definition of $\rho$ in
\eqref{anglepick}, we see that for $\rho=0$ the ellipse is a
circle and for $\rho=\frac{\pi}{2} $ it degenerates to a line. For
generic values of $\rho$, the $SO(3)$-orbit is $SO(3)/\ZZ_2$, with
the $\ZZ_2$ generated by the 180$^\circ$-rotation about the 1-axis,
but for $\rho=0$ the orbit is a 2-sphere and for $\rho=\frac{\pi}{2}$
the orbit is $\RP^2$. This is the same as the orbit structure of AH
compactified by an $\RP^2$ at infinity, although the metric is of
course different.

We also note that the K\"ahler form 
\bee
\omega = i  \frac{|\bz^2| d \bz^t  \wedge d\bar{\bz} - \bz^\dagger 
d\bz \wedge \bz^td\bar{\bz}}{|\bz|^4}
\eee
takes the simple form
\bee
\label{cp2harmonic}
\omega=  \cos\rho\;  \eta_2\wedge\eta_3-\sin\rho \; d\rho \wedge
\eta_1 \,,
\eee
which should be compared to the expression given in \cite{GibbonsPope}  
in coordinates adapted to the $U(2)$ symmetry of $\CP^2$. The form 
$\omega$ is invariant under the 180$^\circ$-rotation about the 1-axis  
and hence well-defined on the generic $SO(3)$-orbits. It is manifestly 
closed, but not exact: for $\rho >0$ we can write $\omega = 
d(\cos \rho \, \eta_1)$ but this  expression is not valid on the 
exceptional $SO(3)$-orbit where $\rho=0$, since $\eta_1$ is not 
well-defined there. 

The K\"ahler form is self-dual with respect to the complex orientation
(with volume element $dV=\omega^2 =-\sin (2\rho)\, d\rho \wedge
\eta_1\wedge\eta_2 \wedge \eta_3$). Since it is closed, it is also
harmonic.  The existence of a non-exact harmonic, self-dual form on
$\CP^2$ follows from the fact that the signature of $\CP^2$ is 1.
In view of our interpretation of signature as baryonic charge one
might expect there to be a  baryonic interpretation of $\omega$.  We
return to this question when discussing the AH model of the proton
further in Sect.~\ref{enfluxsection}.

\subsection{Hitchin's one-parameter family  of Einstein metrics}
\label{Hitchinsection}

If the $\CP^2$-model for the neutron turns out to be too na\"ive, there
is a more sophisticated variant which could be explored. This arises
from the sequence of self-dual Einstein manifolds $M(k)$, for $k>2$,
studied by Hitchin \cite{Hitchin}. The manifolds $M(k)$ for even
$k\geq 4$ are all defined on the same space as our proton model,
namely $\CP^2$ but with $\RP^2$ removed. $M(4)$ is  $\CP^2$  with
the Fubini-Study metric and all  the metrics on $M(k)$, $k>4$ and
even, are incomplete on the open set in  $\CP^2 \setminus \RP^2$, but
can be completed to metrics on $\CP^2$ with a conical singularity of
angle $4\pi/(k-2)$ along $\RP^2$. For odd $k\geq 3$ the manifolds are 
defined on $S^4$, with $M(3)$ being $S^4$ with its standard metric. 
This time the metrics of $M(k)$, $k>3$ and odd, are
incomplete on the open set in $S^4 \setminus \RP^2$, but
can be completed to metrics on $S^4$ with a conical singularity of
angle $2\pi/(k-2)$ along $\RP^2$. 

The sequence of conical manifolds $M(k)$ for even $k\geq 4$  
and starting with $\CP^2$, has decreasing scalar curvature and converges to AH
as $k\rightarrow \infty$. It may turn out that some other
value of $k$ gives a better model for the neutron than $k=4$. 
Note that for $k>4$ the conical singularity breaks the symmetry down
to $SO(3)$. Even for $k=4$ we shall see later that other factors
break the symmetry in this way.

Hitchin also pointed out \cite{Hitchin,Hitchin2} that the family 
$M(k)$ can be extended to  real parameter values.
For any real $k>0$, $M(k)$ is related to the moduli space of 
centred $SU(2)$ monopoles over hyperbolic space of curvature $-1/ p^2$, where
$p = (k-4)/4$. When $k$ is not an integer, the conical angle is not a 
rational multiple of $\pi$ and $M(k)$ is not an
orbifold. Consequently, the explicit methods of \cite{Hitchin} do
not then apply. Nonetheless, having $k$ as a real parameter gives 
useful room for manoeuvre in modelling the neutron and may provide 
contact with conventional nuclear models. In particular, $1/k$ may
play a role as a small parameter that controls the breaking of isospin 
symmetry.

The forthcoming paper  \cite{AtiyahLeBrun} contains a  signature formula for  
Riemannian manifolds with conical singularities, like the Hitchin manifolds $M(k)$. We  summarise that result in 
Sect.~\ref{sigsect}. 
 
\section{The neutrino}

Having put forward models of the electron, proton and neutron, it is
then natural to look for a similar model of the neutrino. Since it
has no electric charge it should, like the neutron, be modelled by a compact
manifold. It should have symmetry similar to the $U(2)$ symmetry of 
the electron and should have positive curvature. It should have 
zero baryon number, that is, vanishing signature. 
 
Just as $\CP^2$ is the most obvious model for the neutron, the standard 
4-sphere, $S^4$, is the most obvious model for the neutrino. Again this has
more symmetry than we need, $SO(5)$ instead of $SO(3)$. Just as a
distinguished $\RP^2$ in $\CP^2$ picks out the smaller $SO(3)$
symmetry, so a distinguished $S^2$ in $S^4$ is needed to cut down the symmetry
of the neutrino.

To exhibit the  symmetry, we parametrise $S^4$ in terms of vectors 
$\vec{x}\in \RR^3$ and $\vec{y}\in \RR^2$ satisfying the constraint
\bee
\vec{x}\cd\vec{x} + \vec{y}\cd \vec{y} =1 \,. 
\eee
The metric is then
\bee
ds^2= d\vec{x}\cd d \vec{x} + d \vec{y}\cd d  \vec{y} \,.
\eee
The group $SO(3)\times SO(2)$ acts in the obvious way on the pair of 
vectors $(\vec{x},\vec{y})\in S^4$ and preserves the metric. In order 
to compare with other metrics discussed in this paper we parametrise
\begin{align}
\vec{x} & = \sin \rho \, (\sin\theta \cos \phi, \sin\theta \sin \phi,
\cos \theta) \,, \nonumber \\
\vec{y}& = \cos \rho \, (\cos\chi, \sin\chi) \,, 
\end{align}
with $\rho \in [0,\frac{\pi}{2}]$, $\chi \in [0,2\pi)$ and the usual 
range for the polar coordinates $\theta, \phi$ on $S^2$,
and find the  expression 
\bee
\label{s4metric}
ds^2= d\rho^2 + \sin^2 \rho \, (\eta_1^2 + \eta_2^2) + \cos^2 \rho \,
d \chi^2 \,.
\eee
Here we used that $ d\theta ^2 + \sin^2\theta \, d\phi^2  
= \eta_1^2 + \eta_2^2$ in terms of the left-invariant 1-forms   
defined in \eqref{linv}. The generic $SO(3)\times SO(2)$ orbit 
is $S^2\times S^1$, but this collapses to $S^1$ when $\rho=0$ and 
to $S^2$ when $\rho=\pi/2$.

Note that $S^4$ is conformally flat, so the Weyl tensor vanishes and
is trivially self-dual. $S^4$ has no middle-dimensional homology, so
the signature is zero, and hence our model neutrino has zero baryon
number, as required. Since $S^4$ also has an orientation-reversing
isometry, our model seems to suggest that the neutrino coincides
with the anti-neutrino. For this and other reasons (see later) our
choice of $S^4$ is very tentative and provisional. As with the $\CP^2$  model
for the neutron it should be regarded at present as a prototype.

To address the symmetry breaking issue, and several others, we will in
the next section discuss how our various models are supposed to fit
into conventional 3-space. 

\section{Attaching the models to space}

So far our models are abstract objects, 4-manifolds on their own,
which are supposed to model four basic particles of nature.  How
are we to view them in the real world?

Let us begin with the easiest case, that of the electron.  Thought of
originally as the Dirac monopole,  the idea is well-known.  We consider
Kaluza-Klein space as a Riemannian 4-manifold with a circle action.
Away from matter this space is assumed to be a circle bundle over
$\RR^3$. Outside a given region in $\RR^3$ which is electrically neutral the
bundle is assumed to be (topologically) just the product space.  If a
region is electrically charged the circle bundle over the boundary is
supposed to have a Chern class equal to minus the charge.  If we start from
the vacuum, then inserting one electron amounts to attaching a
truncated version of TN to the boundary. This truncation turns
the idealized model into a more realistic model of a particle. If 
other particles are
present, TN will be an approximation to the precise metric. This 
approximation is some measure of the force exerted on the electron. 
In a dynamic theory, forces should emerge from the equations, a task 
for the future.

Next let us move on to the proton. This is similar, using a truncated
version of AH. However, as pointed out earlier, the circle fibration
is now not oriented so that the asymptotic 3-space is not $\RR^3$
(unless our region contains equal numbers of protons and anti-protons).

The model for the neutron  is compact, so there is no way to attach it
to a boundary. Instead we propose that our model neutron (a copy of
$\CP^2$) intersects our 4-space in a surface. This surface should
project to a surface in 3-space which is the `boundary' of the neutron
as seen by an observer. Since we want to keep the neutron similar to
the proton (except for the charge) it seems reasonable to take this
surface to be a copy of $\RP^2$. But since the charge is now zero the
circle bundle over the surface should be trivial. This could arise as
follows. Pick a point in $\RR^3$ and blow it up to give an $\RP^2$,
so that we are modifying 3-space in this neighbourhood. Keep the
circle bundle trivial, so making the charge zero. Lift this
$\RP^2$ into the total space of the circle bundle and let the $\CP^2$ neutron
model intersect 4-space in this $\RP^2$, which is the
distinguished $\RP^2$ in $\CP^2$ (hence breaking the larger symmetry
of $\CP^2$). In this construction the metric on $\CP^2$ need not be 
changed. The only change that is needed is the change in metric on the 
background 4-space got from the blowing up process in 3-space.

Finally we come to the neutrino. From the other examples it is clear
what is required. This time we just take a 3-ball in $\RR^3$ with boundary
a 2-sphere. The circle bundle over it is trivial and we lift the sphere
to the total 4-space. We now require the $S^4$ neutrino model to
intersect our 4-space in the chosen 2-sphere. The choice of the
2-sphere in $S^4$ again breaks the symmetry, down to $SO(3)\times SO(2)$.
The 2-sphere need not be a great (geodesic) sphere and this provides 
a parameter to play with.

Note that in this case we could reinterpret the picture as the surgery
that kills off the circle and leaves a 2-sphere. This means that for
the electron, the proton and the neutrino we can still think of our
`space' as a 4-manifold. But this does not seem to work for the
neutron where we have to settle for a 4-space with intersecting 
components like a complex algebraic surface with double curves.

The four models that we have proposed for the four basic particles
should be geometrically related in some way to account for the process
of beta decay in which a neutron breaks up into a proton, an electron
and an anti-neutrino. The opposite asymptotic behaviour of AH and TN is a
good start but the difference in the asymptotic fundamental groups
presents a problem. This suggests that the model of the neutrino
should somehow bridge the gap and it argues against the simplicity of
the 4-sphere. We hope to pursue this question.

\section{Spin 1/2}
\label{spinsect}

In all our models we have a natural action of the symmetry group of rotations
($SO(3)$ or $SU(2)$) preserving the metric and the `asymptotics',
the details of which differ according to the cases. For
the neutral, compact models we interpret `asymptotic' to mean the
behaviour near the distinguished surface where the 4-manifold
intersects 3-space, which is either an $\RP^2$ or an $S^2$. For the
electrically charged, non-compact models we have an asymptotic
fibration by circles over physical 3-space; rotations preserve this
fibration and induce an $SO(3)$-action on the base. The
hyperk\"ahler structures on TN \cite{GibbonsRuback} and AH
\cite{Olivier}  can be used to construct  Cartesian coordinates on physical 3-space 
with the physically required transformation properties under spatial rotations.
Here we make essential use of the fact that the complex structures on TN and AH 
transform as  vectors   under rotations, as explained at the beginning of Sect.~\ref{protonelectron}.

For a model to represent a particle of spin 1/2 we must include the data 
necessary to lift the rotation group action to an $SU(2)$ action, and 
to construct its 2-dimensional representation. To achieve this for non-compact
(electrically charged) models we require a spin structure on the
4-manifold while for compact (neutral) models we only require a spin
structure on the inside,  obtained by removing a distinguished surface
$X$ from the 4-manifold. In this short section we explain 
 why, in each of the models considered in this paper, 
 the lift of the rotation group action to  the spin bundle is an $SU(2)$-action.
 We  do not attempt to construct naturally associated spin 1/2 
representations, but comment on how this may be done.

For the TN model of the electron, there is nothing to do since the 
rotation group action on TN is an $SU(2)$-action. 
For the neutron  model we view the required data as the
compactification of the proton model AH, not just topologically, but
also with the action of the rotation group. In particular the $\RP^2$
at infinity is part of the data. We now simply require the extra data
of a spin structure on the inside $\CP^2\setminus \RP^2$, i.e., on AH.

It might be thought that, since the spin structure on AH is unique,
there is nothing gained by the additional data, but this is to ignore
the interaction with the symmetries. We must now require the rotation
group to lift to the spin bundle, and this may require us to
pass to $SU(2)$, in which case we label the model as fermionic.
Otherwise we call it bosonic. To see that, in principle, either case 
could occur, consider the two manifolds $Y_1 = SO(3) \times  \RR$ and 
$Y_2 = \RR^3 \times  \RR$.  We take the left-translation action of $SO(3)$ 
on $Y_1$ and the standard action on $Y_2$. For $Y_1$, the tangent bundle
is trivial and we can choose the trivial spin bundle (though since
$Y_1$ is not simply-connected there is another choice). The
$SO(3)$-action extends without going to $SU(2)$, so $Y_1$ is bosonic
in our terminology. For $Y_2$ the fixed point at the origin of $\RR^3$
means that we can only lift to the spin bundle after passing to
$SU(2)$, so $Y_2$ is fermionic.
 
For AH we have to show that, with its action of $SO(3)$, it is
fermionic. There are several ways to do this. Perhaps the simplest
(in line with the example $Y_1$ above) is to note that the action is
not free and that the isotropy group of a point on the core is
$SO(2)$. To lift even this subgroup to the spin bundle over the fixed
point requires us to go to the double cover.

A comparison between our models for the electron and proton is
illuminating. As we pointed out, $SU(2)$ acts on the electron but only
$SO(3)$ acts on the proton. Thus the fermionic natures of our two
models differ. In one case it is inherent in the symmetry while in
the other it is geometric or topological.  

Our tentative model for the neutrino is just the round 4-sphere, with
a distinguished 2-sphere at infinity given by the decomposition
$\RR^5  = \RR^3 \times  \RR^2$  and the corresponding action of
$SO(3)$. The inside, got by removing $S^2$, has an infinite cyclic
fundamental group so there are infinitely many spin structures, but
only one extends to the whole of $S^4$. If we pick this, then it is
easy to see that the lift of $SO(3)$ to the spin structure requires us
to pass to $SU(2)$ (for example we can use an $SO(2)$ fixing a point
at infinity and argue as with the proton). Thus our model of the
neutrino is fermionic.

Our discussion so far shows that our geometric approach furnishes
fermionic models,  but it does not establish that they necessarily
give spin 1/2. This requires constructing the 2-dimensional
representation of $SU(2)$ and relating it to the asymptotic region. We
expect that the required 2-dimensional representations can be
constructed in terms of eigenspaces of the Dirac operator on our
model manifolds, possibly twisted by a $U(1)$-bundle with curvature
proportional to one of the harmonic 2-forms discussed in
Sects.~\ref{cp2sect} and \ref{enfluxsection}. 
 
\section{Charges, energies and fluxes}
\label{enfluxsection}
\subsection{General remarks}
\label{sigdef}
So far we have focussed on topological and geometrical features of 
our models and explained how they describe general properties of 
particles -- like  baryon number, electric charge and location in space.
We want to keep an open mind about how our geometric models make
quantitative contact with the physics of elementary particles. In
particular, we do not assume that this should necessarily happen in
the standard framework of Lagrangian field theory, where dynamics,
conservation laws and even the quantum theory are all derived from
an action functional. 

The purpose of this section is to illustrate that our geometric models 
for particles nevertheless contain natural candidates for the kind of 
quantities which arise in Lagrangian models, like energy density and 
electric fields. We show that electric charge as defined in our model can
be represented by a harmonic 2-form, thus making contact with the
usual description of electric flux. One important feature
of the densities and fields considered in this section is that they
are defined on the 4-manifold so that they can only be interpreted
as conventional {\em spatial} densities and fields in the asymptotic
region of the 4-manifold which canonically projects down to physical
3-space.
 
We begin with a summary of the topological quantities and their
physical interpretation for each of the 4-manifolds considered thus
far. For compact manifolds the electric charge is zero and for
non-compact manifolds it is minus the  self-intersection number $X^2$
of the manifold $X$ representing infinity in their compactification. The
compactification is $\CP^2$ for both TN and AH, but $X=\CP^1$ for TN
with self-intersection number $X^2=1$, while $X=\RP^2$ for AH, with
self-intersection number $X^2=-1$; see Appendix \ref{signsTNAH} for
details.

The baryon number is identified with the signature of the
4-manifold. 
 For a non-compact
oriented manifold, the signature is defined as the signature of the
image of the compactly supported cohomology in the full cohomology
\cite{AtiyahPatodiSinger}.   The topology of TN is that of $\RR^4$ so the signature vanishes. 
The signature of AH is 1. This  follows from the fact that its 2-dimensional homology is generated by the core 2-sphere, and that the self-intersection number of this 2-sphere is positive (in fact equal to +4).  The same argument applies  to the  sequence of Hitchin manifolds $M(k)$  for $k\geq 4$ and even, reviewed in
Sect.~\ref{Hitchinsection}, which are all compact  and topologically equivalent to $\CP^2$.

In Table~\ref{manifoldlist} we list the electric charge and baryon number  as well as the Euler 
characteristic $\chi$ and the squared $L^2$-norm $||R||^2$ of the Riemann
curvature for  the four 4-manifolds mainly discussed in this paper.
The Euler characteristic is homotopy
invariant, so can be computed for TN and AH by noting that the former
retracts to a point and the latter to a 2-sphere.

\begin{table}[h!]
\begin{center}
\begin{tabular}{|c|c|c|c|c|}
\hline 
       & $-X^2$ (electric charge)  &  $\tau$ (baryon number)  
& \quad $\chi$ \quad &  \quad  $||R||^2/(8\pi^2)$  \quad   \\
       \hline 
TN (electron)     &     $-1$      &        0  
& \quad $ 1 $  \quad     &  \quad  $1$  \quad   \\
\hline 
AH (proton)     &       $ \phantom{-} 1$     &    1   
&  \quad $2$  \quad    & \quad   $2$ \quad   \\
\hline 
$\CP^2$ (neutron) &     $ \phantom{-} 0$    &     1     
&  \quad $3$  \quad  &  \quad  $3$  \quad    \\
\hline 
$S^4$ (neutrino) &    $ \phantom{-} 0 $     &   0     
& \quad $ 2$  \quad   &  \quad $2$   \quad  \\
\hline 
\end{tabular}
\caption{Geometric properties of 4-manifolds and their physical interpretation}
\label{manifoldlist}
\end{center}
\end{table}

\subsection{Signature, Euler characteristic and $L^2$-norm of the 
Riemann curvature}
\label{sigsect}

For compact Riemannian 4-manifolds there are   formulae for the Euler characteristic and signature 
 in terms of integrals over the
4-manifold involving the Riemann curvature, which we shall review
below.   In the non-compact cases,  these bulk contributions need to be
supplemented by boundary integrals  and
(for the signature) a subtle spectral contribution ($\eta$-invariant), see e.g.  
\cite{EGH} and \cite{AtiyahLeBrun} for a summary.
For   manifolds with conical singularities like the sequence of Hitchin manifolds  reviewed in
Sect.~\ref{Hitchinsection},  a signature formula was recently found
 \cite{AtiyahLeBrun},  which we also 
review   below. We now discuss the bulk contributions, relegating 
most detailed calculations to Appendix \ref{Riemanncalculations}.

Writing $R$ for the Riemann tensor as in Appendix
\ref{AHTNdetails}, we define the squared $L^2$-norm of $R$ (for
compact and non-compact manifolds) as
\bee
\label{l2defined}
||R||^2 = \int_M \sum_{a<b} R_{ab}\wedge *R_{ab} \,,
\eee
where $*R_{ab}$ is the Hodge dual of $R_{ab}$.
The form that integrates to the first Pontrjagin class on a compact 
manifold is 
\begin{align}
\label{Pontrjagin}
p_1 =  \frac{1}{4\pi^2} \sum_{a<b} R_{ab}\wedge R_{ab} \,,
\end{align}
and the form which integrates to the signature $\tau$ in the compact case is 
\bee
S=\frac{1}{3} p_1 \,,
\eee
so that 
\begin{align}
\tau(M) =\int_M S= \frac 1 3 \int_M p_1 = 
\frac{1}{ 12\pi^2 }\int_M \sum_{a<b} R_{ab}\wedge R_{ab} \,.
\end{align}
The form that integrates to the Euler characteristic $\chi$ in the 
compact case is  
\begin{align}
e=  \frac{1}{16\pi^2} \sum_{a<b} \epsilon^{abcd} R_{ab}\wedge R_{cd} \,.
\end{align}
Note that, with our conventions and for compact Einstein manifolds 
\cite{LeBrun},
\bee
\chi(M) = \int_M e = \frac{1}{8\pi^2} ||R||^2 \,,
\eee
which determines $||R||^2$ for $\CP^2$ and $S^4$ in terms of their topology.

The Riemann curvature on a 4-manifold may also be thought of as a
mapping of 2-forms. Exploiting the fact that the space $\Lambda^2$ of
2-forms on an oriented 4-manifold decomposes into $\pm$-eigenspaces
$\Lambda^\pm$ of the Hodge star operator $*$, we get a corresponding
decomposition of the Riemann curvature into irreducible pieces \cite{EGH} 
\bee
R = \begin{pmatrix} W^+ +\frac{s}{12} & B \\ *B &  W^-+\frac{s}{12}  
\end{pmatrix} \,.
\eee
Here $W^\pm$ are the self-dual and
anti-self-dual parts of the Weyl tensor, $s$ is the scalar curvature
and $B$ amounts to the tracefree part of the Ricci curvature. Then the
signature of a compact manifold can also be expressed as \cite{LeBrun}
\bee
\tau(M) = \frac {1}{12\pi^2} \left(||W^+||^2 -  ||W^-||^2 \right).
\eee
For a self-dual manifold $W^- = 0$ so, up to the factor $12\pi^2$,
$\tau$ is given by the $L^2$-norm of the Weyl tensor $W^+$, and is
non-negative. 
 
Since the metrics on both TN and AH are hyperk\"ahler, $B$ and $s$
vanish, so the full Riemann curvature is self-dual. Therefore the 
bulk contribution to both the signature and the Euler characteristic 
can be expressed in terms of the $L^2$-norm of the Riemann curvature,
which equals the $L^2$-norm of $W^+$. 
In Appendix \ref{Riemanncalculations}, we show that
\bee
||R||^2_{{\rm TN}}= 8\pi^2 \,, \qquad ||R||^2_{{\rm AH}} =16\pi^2 \,.
\eee
Using the self-duality of the Riemann curvature, we deduce that the
bulk contributions to the Euler characteristic are both in agreement
with the topological results listed above:
\bee
\int_{\rm TN}e =\frac{8\pi^2}{8\pi^2} =1 \,, \quad 
\int_{\rm AH} e=\frac{16\pi^2}{8\pi^2} =2 \,. 
\eee
The bulk contributions to the signatures, on the other hand, turn out 
to be fractional:
\bee
\label{bulkAHTN}
\int_{\rm TN} S=\frac{8\pi^2}{12\pi^2} =\frac{2}{3} \,, 
\quad \int_{\rm AH} S =\frac{16\pi^2}{12\pi^2} =\frac{4}{3} \,.
\eee
As shown in \cite{AtiyahLeBrun},  the fall-off of the spin connection and curvature imply  that the boundary integrals do not contribute in the limit to either the Euler characteristic or the signature. However,  the fractional values of the bulk integrals for the signature show that there must be a non-zero contribution from the $\eta$-invariant.

In \cite{AtiyahLeBrun}, a  signature formula is derived for  Riemannian 4-manifolds with  conical singular metrics.   As reviewed in Sect.~\ref{Hitchinsection},  the Hitchin manifolds $M(k)$ are of this type, and $M(\infty)$ can be identified with AH. The
general formula derived in \cite{AtiyahLeBrun} for the signature of  a
Riemannian 4-manifold $M$ whose metric has a conical singularity of
angle $2\pi/\kappa$ along a surface $X$ is
\bee
\tau(M)=\frac{1}{3}\int_M p_1 - \frac 1 3
\left(1-\frac{1}{\kappa^2}\right)X^2 \,,
\eee
where $p_1$ is the form \eqref{Pontrjagin} which integrates to the
Pontrjagin class, and $X^2$ is the self-intersection number of $X$ in
$M$. 

\begin{figure}[!ht]
\centering
\includegraphics[width=7truecm]{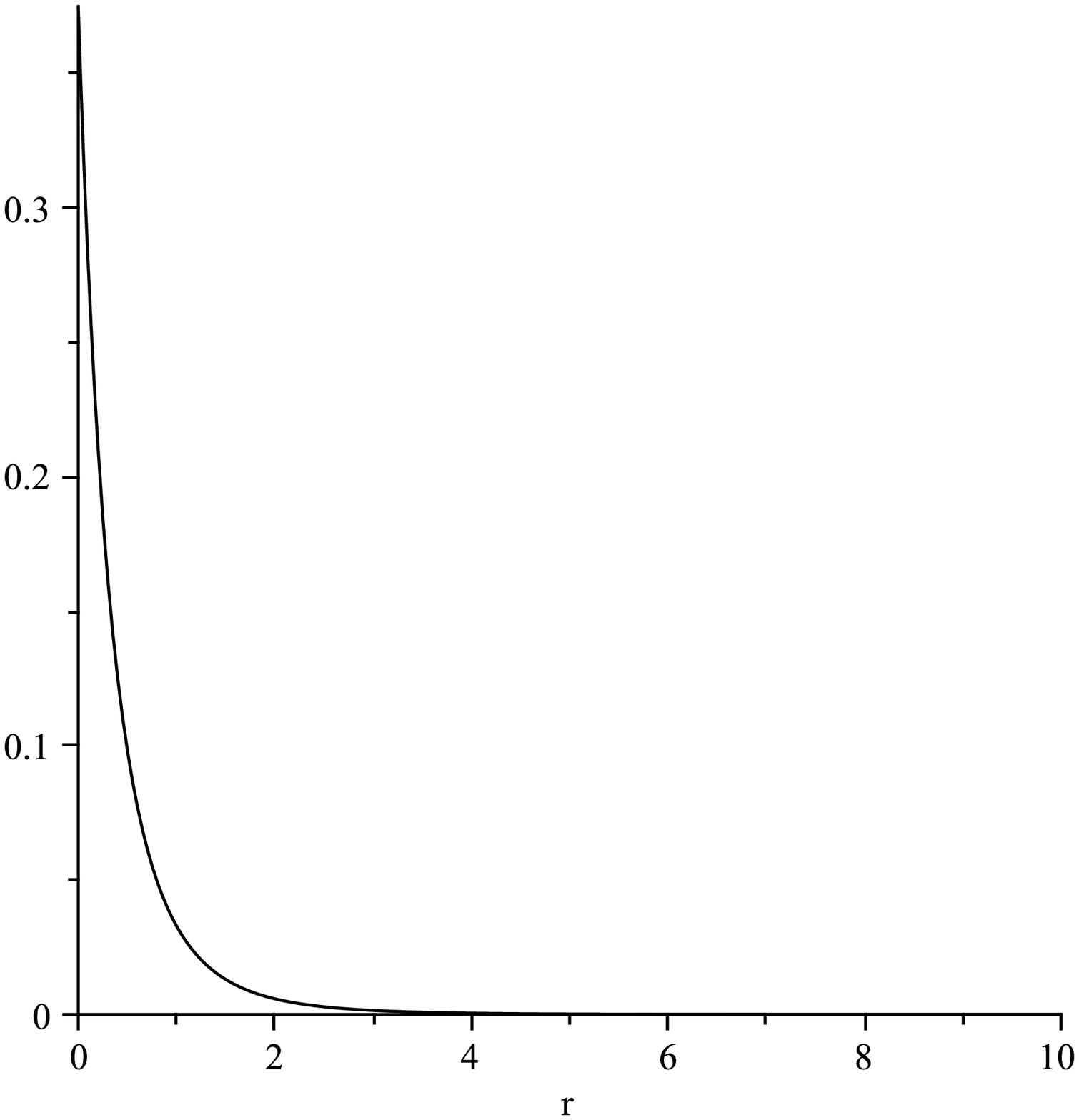}
\includegraphics[width=7truecm]{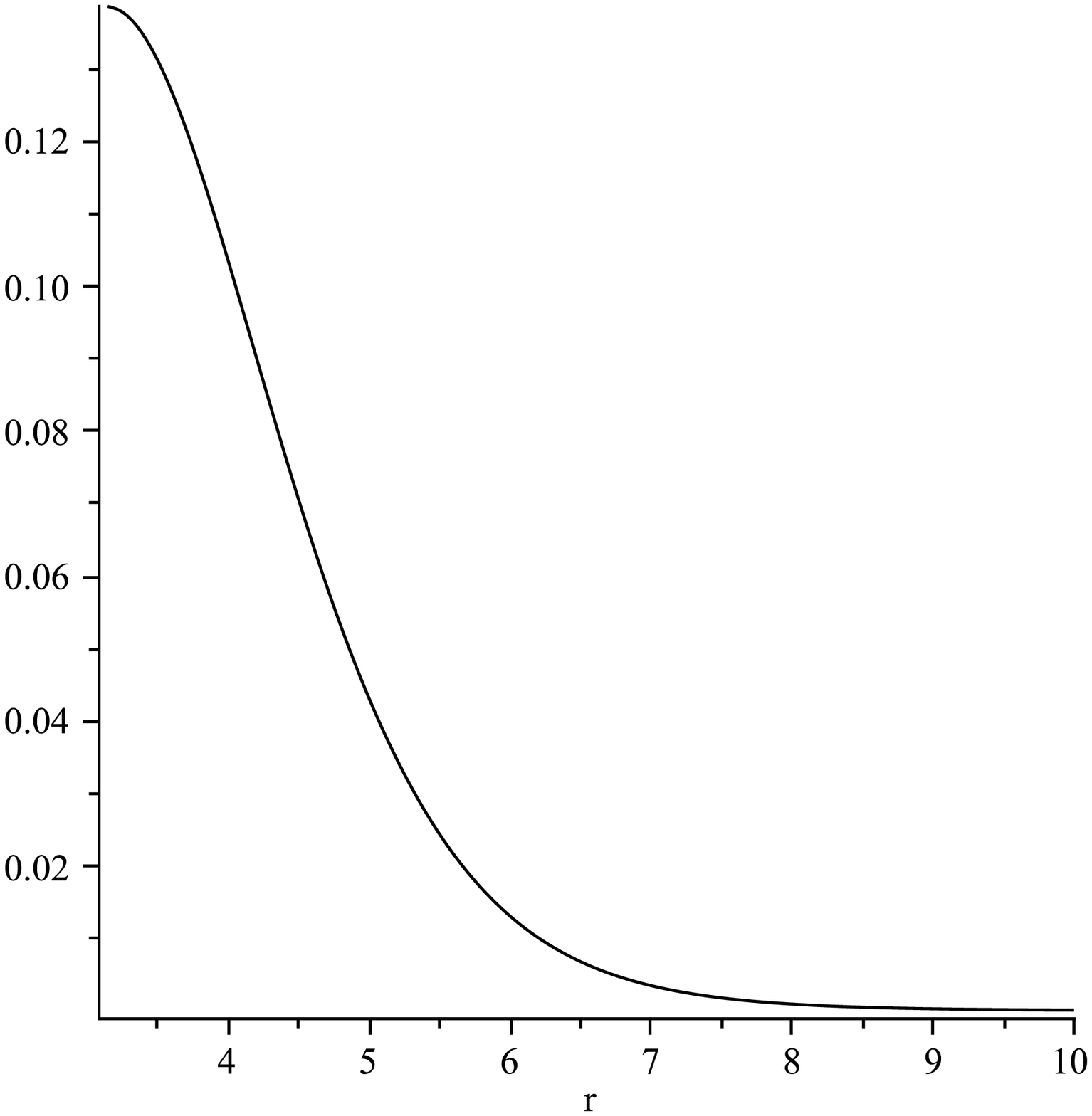}
\caption{The density \eqref{curvkernel} for 
TN (left) and AH (right)}
\label{curvdens}
\end{figure}

Since the signature $\tau$ is interpreted as the baryon number in our
model, it is tempting to interpret the integrand of the bulk
contribution to the signature as a baryon number density. Our
calculations above show that this cannot be the whole story, since the
signature $\tau$ also receives a contribution from the
$\eta$-invariant of the boundary. We nevertheless compute and plot
the integrands of the bulk contributions to the signature below. Since
overall factors are not important here, we look at the squared
$L^2$-norm \eqref{l2defined} of the Riemann curvature.  In the
notation of Appendix \ref{AHTNdetails}, the integrand of
\eqref{l2defined} in both the TN and AH case can be written as 
\bee
dF\wedge\eta_1\wedge\eta_2\wedge\eta_3 = \frac{r}{ab^2c}\frac{dF}{dr} \, 
\d V \,, 
\eee
where $\d V$ is the volume element, so that the combination  
\bee
\label{curvkernel}
\frac{r}{ab^2c}\frac{dF}{dr}
\eee
may be interpreted as a density. We plot this density for TN and AH in
Fig.~\ref{curvdens}. It is finite at, respectively, the origin
and the core, and for large $r$ falls off like $1/r^6$. In the
AH case we compute the functions $a,b,c$ and hence $F$ numerically;
in the TN case we have the explicit formula
\bee
\frac{r}{ab^2c}\frac{dF}{dr}=\frac{24}{(r+2)^6} \,.
\eee

\subsection{Electric and baryonic fluxes}

We now construct fields on TN and AH which carry the electric flux
measured by the self-intersection numbers $X^2$ tabulated in
Table~\ref{manifoldlist}. We do this by systematically studying
rotationally symmetric harmonic forms on both TN and AH, with most
of the details given in Appendix~\ref{harmonicsect}.

One finds \cite{Brill,Pope,LeeWeinbergYi,GauntlettLowe} that the 
only square-integrable harmonic and rotationally symmetric 2-form on 
TN is, up to an overall arbitrary constant,
\bee
\label{TNelectricfield}
\omega_3^+ =\left( \frac{r}{r+2} \, \eta_1\wedge \eta_2  
+ \frac{2}{(r+2)^2} \, dr\wedge \eta_3\right) 
=  d\left(\frac {r}{r+2} \, \eta_3\right) \,.
\eee
We claim that this form is a harmonic representative of the Poincar\'e
dual of the surface  $X=\CP^1$ at infinity in TN, the surface
parametrised by $\theta$ and $\phi$. Our calculation
also gives an alternative computation of the electric charge as
minus the self-intersection number of $\CP^1$ in $\CP^2$.
Let $p$ be this self-intersection number, then 
\bee
\omega_{\CP^1} = -\frac{p}{4\pi} \, \omega_3^+
\eee
is Poincar\'e dual to $\CP^1$  since 
\bee
\int_{\CP^1} \omega_{\CP^1} = p \,.
\eee
However, by the definition of the self-intersection numbers in terms
of cohomology, we also have
\bee
\label{electronintersect}
\int_{{\rm TN}} \omega_{\CP^1}\wedge\omega_{\CP^1} = p \,.
\eee
Evaluating the integral in Appendix~\ref{harmonicsect} we find
\bee
p^2 =p \,,
\eee
so that $p=1$, confirming that, with the self-dual orientation, the
self-intersection of $\CP^1$ in $\CP^2$ is 1. 

Since the 2-form $-\omega_{\CP^1}$ is harmonic and since its total
flux through infinity equals the electric charge we interpret it
as the electric field of the electron. Although we have adopted a  
viewpoint dual to standard electromagnetism, with a purely 
spatial 2-form being interpreted as electric rather than magnetic
flux, it is interesting that the self-dual 2-form
\eqref{TNelectricfield} also contains a term which allows for a
conventional electric interpretation: when we contract
$\omega_{\CP^1}$ with the vector field $\partial_\psi$ along the
fibres (using $\eta_3(\partial_\psi)=1$),
we obtain a purely radial field which, asymptotically, falls off like $1/r^2$.

\begin{figure}[!ht]
\centering
\includegraphics[width=7truecm]{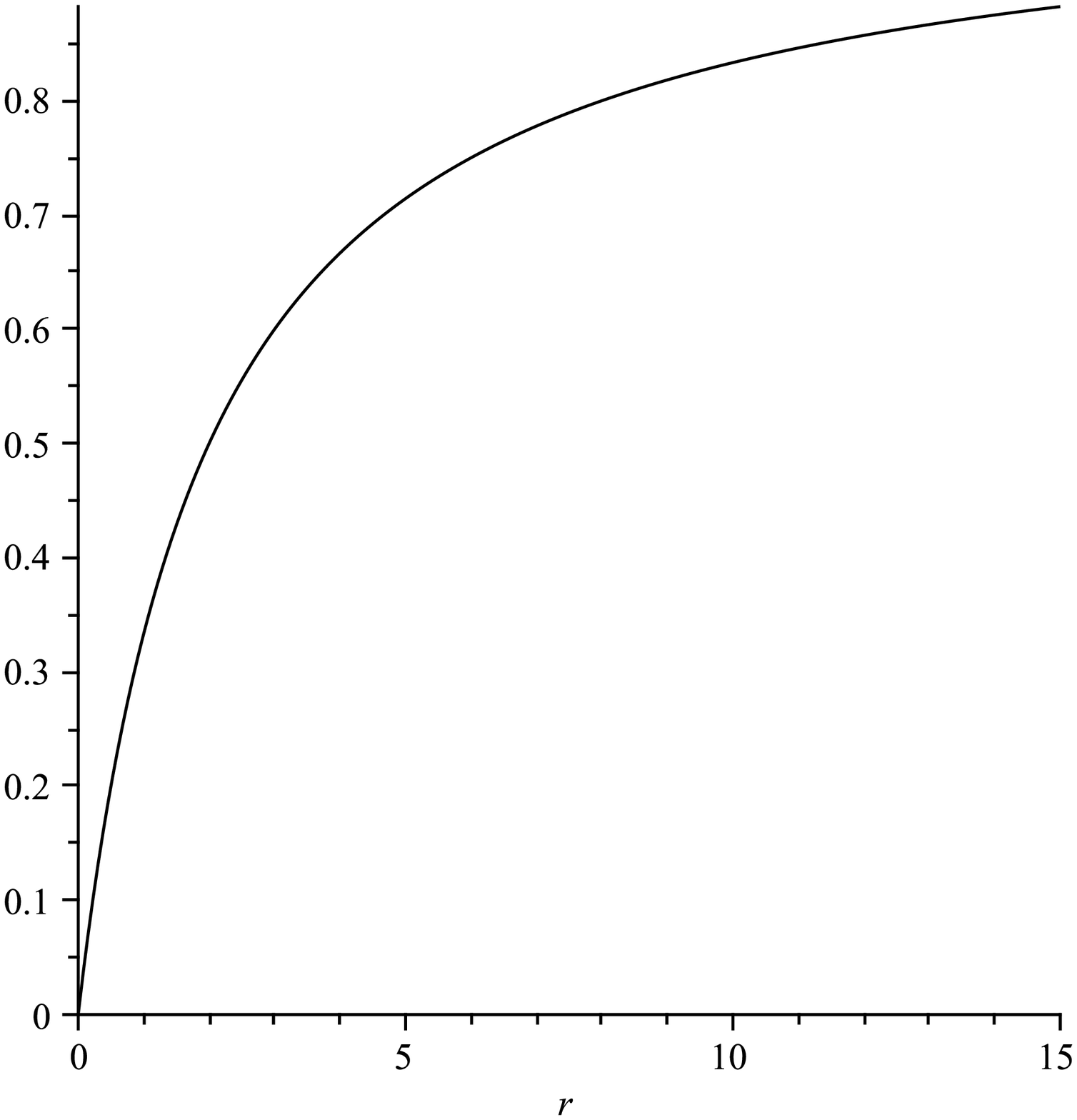}
\includegraphics[width=7truecm]{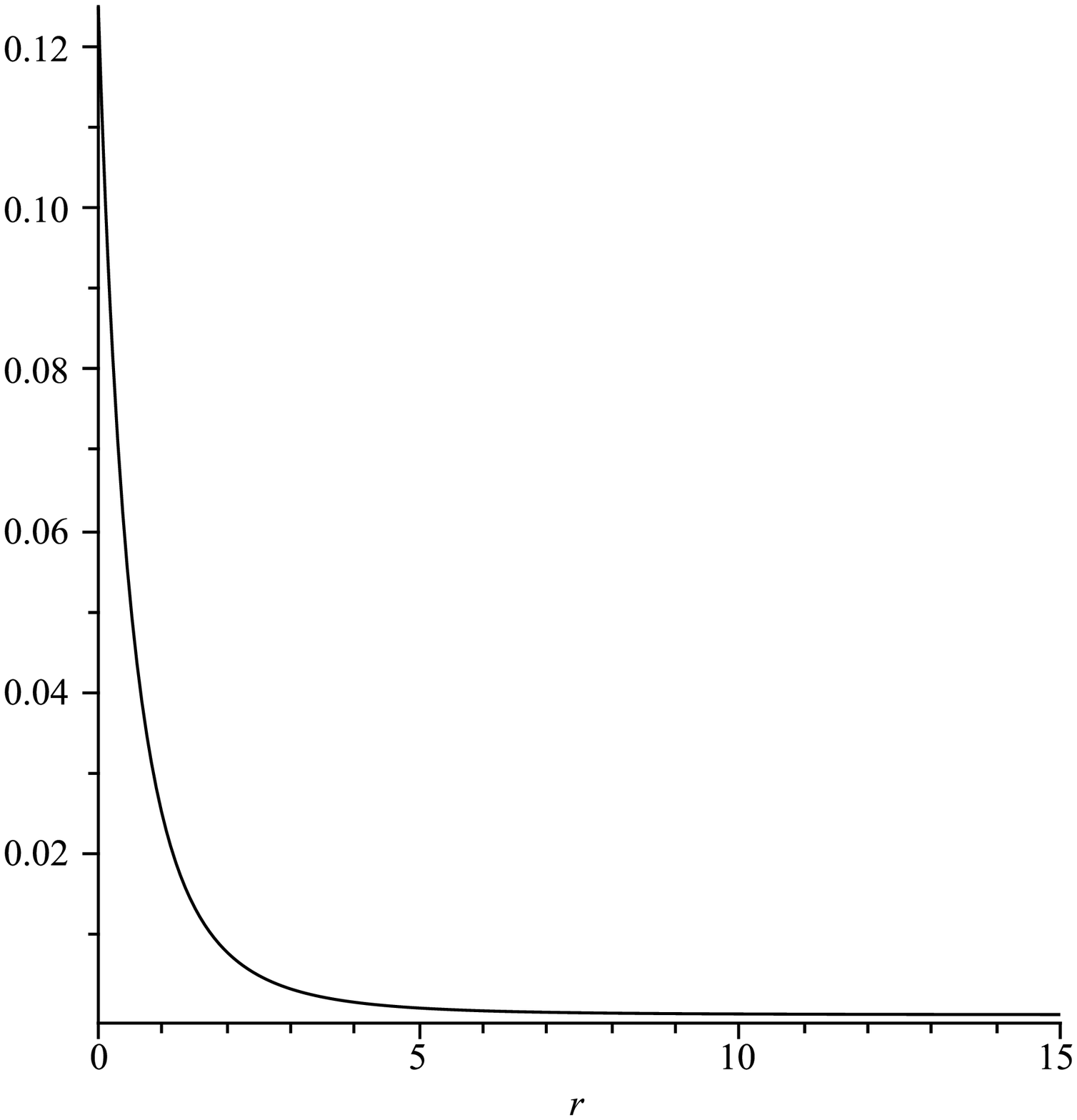}
\caption{The `electric' profile function $g_3^+$ and the energy 
density $\rho_3^+$}
\label{TNenpic}
\end{figure}

The integral \eqref{electronintersect} is the squared
$L^2$-norm of the electric field. Ignoring overall factors and working
with  $ \omega_3^+$ we write 
\bee
||\omega_3^+||^2 = \int \rho_3^+ \, \d V \,,
\eee
with $\d V$ defined as in \eqref{volume}, and interpret the integrand 
\bee
\rho_3^+(r) = \frac{2}{(r+2)^4}
\eee
as an electric energy density. 
For comparison with the AH case below, we plot the profile function
$g_3^+(r)=r/(r+2)$ and the energy density $\rho_3^+$ in Fig.~\ref{TNenpic}.

Turning now to AH, we review in Appendix~\ref{harmonicsect} why there
are only two $SO(3)$-invariant harmonic forms $\Omega^\pm_1$ on
AH which respect the identification \eqref{ident}. They have the structure
\bee
\Omega^{\pm}_1=G^\pm_1  \eta_2\wedge \eta_3  + \d G^\pm_1 \wedge
\eta_1 \,,
\eee
for functions $G^\pm_1$ of $r$ which satisfy   the ordinary differential
equations for $g^\pm_1$ in  \eqref{closed}. Using the 
asymptotic formulae \eqref{AHbolt} and \eqref{AHinfty} one shows 
\cite{GibbonsRuback,MantonSchroers,Sen} that only $G^+_1$ is finite at 
the core and decays at infinity. In fact, the solution decays 
exponentially fast at infinity, with the leading term proportional 
to $e^{-r/2}$. We normalise $G_1^+(\pi)=1$, so that near the core
\bee
\label{harmonicbolt}
G_1^+(r)\sim 1-\frac{1}{\pi^2}(r-\pi)^2 \,, \quad \text{for} \; (r-\pi)\; 
\text{small} \,.
\eee
The 2-form  $\Omega^+_1$ is not dual to the surface $X=\RP^2$ at
infinity in AH and has no interpretation in terms of electric
flux. However, it is dual to the  core  $\CP^1$ in  AH. With
\bee
\Omega_{{\rm core}}= -\frac{p}{4\pi}\Omega_1^+
\eee
one finds
\bee
\int_{{\rm core}} \Omega_{{\rm core}} = p \,.
\eee
As a check, we use this form to compute the self-intersection number 
of the core. Recalling that the volume of $SO(3)/\ZZ_2$ is $-4\pi^2$, and 
using \eqref{masterintAH}, we have 
\bee 
\int_{{\rm AH}} \Omega_{{\rm core}}\wedge \Omega_{{\rm core}} 
=\frac{p^2}{16\pi^2}(-4\pi^2) (0^2 - 1^2) =  \frac {p^2} {4} \,,
\eee
so that
\bee
\frac{p^2}{4} = p \,, 
\eee
confirming $p=4$ for the
self-dual orientation (this agrees with \cite{AtiyahHitchin}, where
the anti-self-dual orientation gives the opposite sign).
  
The  harmonic form \eqref{cp2harmonic} on $\CP^2$ is related to the 
signature of $\CP^2$ through the fact that the signature of a compact 
4-manifold equals the difference of the dimensions of the spaces of 
self-dual and anti-self-dual harmonic representatives of the second  
de Rham cohomology. It seems likely that the (up to scale) unique
bounded and rotationally symmetric self-dual harmonic form 
$\Omega_{{\rm core}}$ on AH is related to the self-dual harmonic form 
\eqref{cp2harmonic} via the sequence of Hitchin manifolds reviewed 
in Sect.~\ref{Hitchinsection}.

\begin{figure}[!ht]
\centering
\includegraphics[width=7truecm]{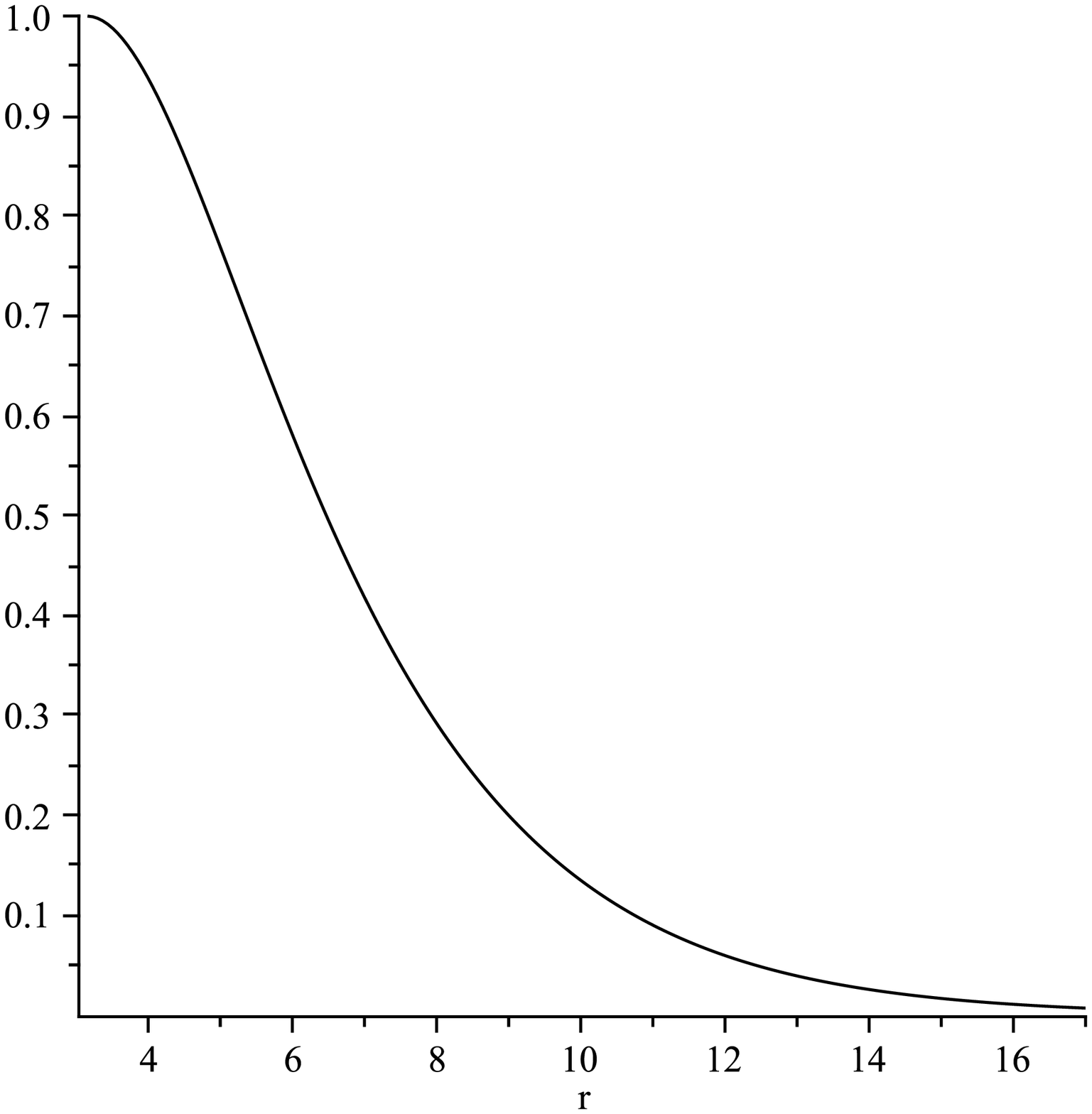}
\includegraphics[width=7truecm]{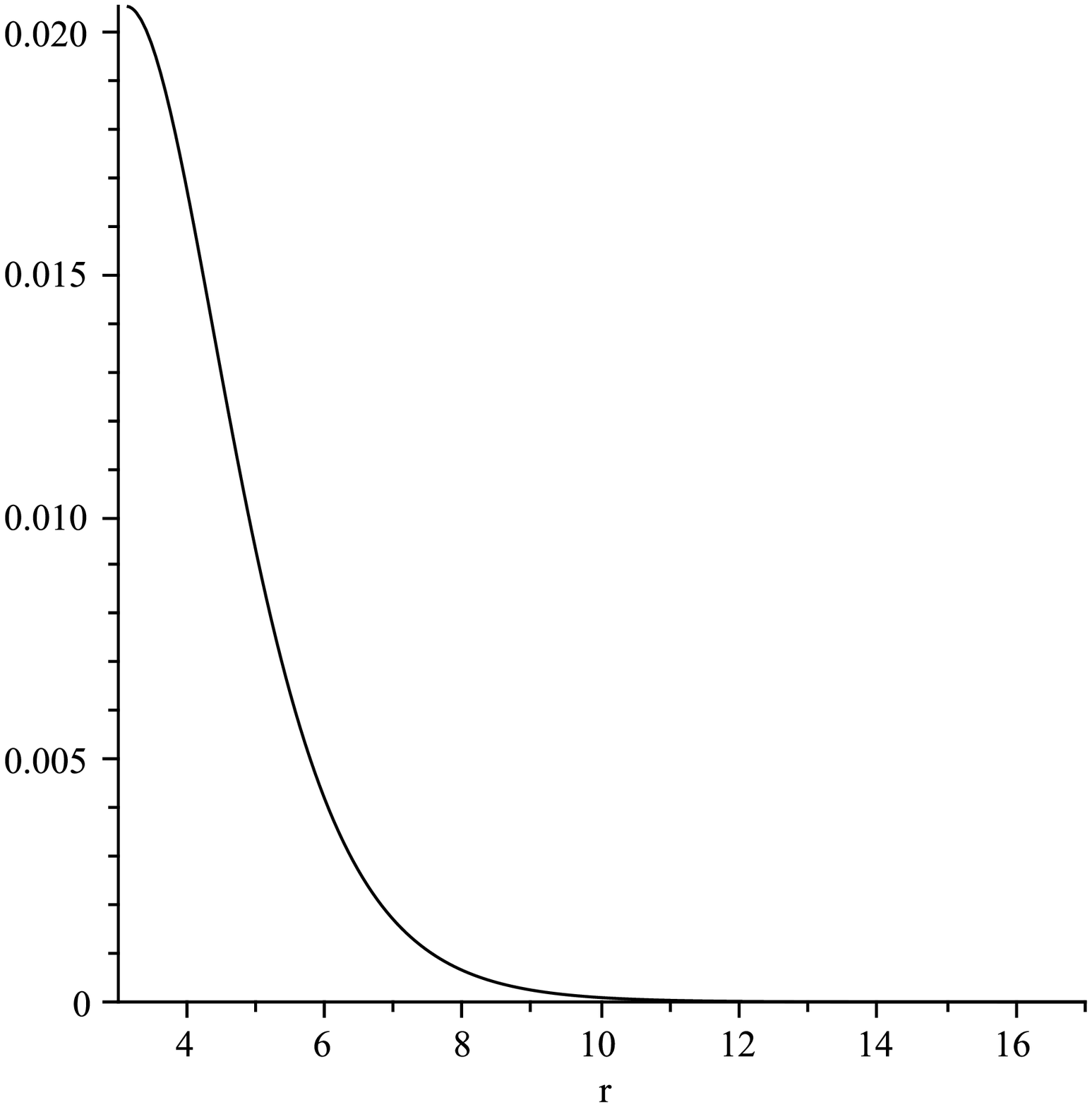}
\caption{The `baryonic' profile function $G_1^+$ and the energy 
density $\rho_1^+$}
\label{AHbaryon}
\end{figure}

Since the existence of $\Omega_{{\rm core}}$ on AH is linked to the 
signature of AH being 1, and since signature represents baryon number 
in our approach, it may be possible to interpret the detailed 
structure of $\Omega_{{\rm core}}$ in baryonic terms. The
exponential fall-off exhibited by $\Omega_{{\rm core}}$ is
reminiscent of the proton's pion field  in the Yukawa description of
the nuclear force. In Fig.~\ref{AHbaryon} we plot the numerically
computed profile function $G_1^+$  and the associated energy density
\eqref{genen}, which turns out to be
\begin{equation}
\rho_1^+= 2\frac{(G_1^+)^2}{b^2 c^2} \,.
\end{equation}
This is finite at the core and decays exponentially according to 
$e^{-r}$ for large $r$. 

To find a harmonic 2-form on AH which can play the role of the
proton's electric field we need to go to a branched cover of AH, 
denoted $\overline{\rm AH}$. The metric on the 
branched cover is not smooth at the core, but this will not affect 
the following calculations near infinity. Dropping the requirement 
that forms are invariant under the identification \eqref{ident}, 
we find that the closure condition on the 2-forms \eqref{formsagainAH} 
has only one further solution which is finite  at the core and which 
remains bounded for large $r$. This is the 2-form
\bee
\Omega^-_3=G^-_3 \eta_1\wedge \eta_2 + \d G^-_3 \wedge \eta_3 \,,
\eee
with a radial function $G_3^-$ satisfying \eqref{closed}. The
solution vanishes at the core as
\begin{equation}
\label{smallg3}
G_3^-(r)\sim C \sqrt{r-\pi} \,, \quad \text{for} \; (r-\pi)\; 
\text{small} \,,
\end{equation}
for some constant $C$.
The large $r$ behaviour is
\begin{equation}
\label{asyg3}
G_3^-(r) \sim \widetilde{C} \, \frac{r-2}{r} \,,
\end{equation}
where $\widetilde{C}$ is another constant. This leads to an
anti-self-dual form on $\overline{\rm AH}$, which is
square-integrable (as we shall show below) and which has not been
considered previously. It is our candidate for the electric
field of the proton. 
 
As a manifold, $\overline{\rm AH}$ compactifies to $S^2 \times S^2$,
as discussed in Appendices~\ref{branched1} and \ref{branched2}. 
The surface $X=\RP^2$ which compactifies AH to $\CP^2$
becomes the  anti-diagonal $S^2$ in $S^2 \times S^2$. Choosing 
$\widetilde{C}=1$ in \eqref{asyg3}, a harmonic representative of the 
Poincar\'e dual class to this $S^2$ is 
\bee
\Omega_{S^2}= -\frac{p}{4\pi} \, \Omega_3^- \,.
\eee
Then 
\bee
\int_{S^2}\Omega_{S^2} = p \,,
\eee
and since the volume of $SO(3)$ is $-8\pi^2$ by
\eqref{angularintegrals}, we also compute
\bee
\int_{\overline{\rm AH}}\Omega_{S^2}\wedge \Omega_{S^2} = 
\frac{p^2}{16\pi^2}(-8\pi^2)(1^2-0^2) = -\frac{p^2}{2} \,.
\eee
Hence
\bee
-\frac{p^2}{2}=p \,,
\eee
which is solved by $p=-2$. Dividing by 2, to take us back to $\RP^2$,
we confirm the self-intersection number of $\RP^2$ in $\CP^2$ as $-1$,
again for the self-dual orientation.
 
Since the 2-form $-\Omega_{S^2}$ on $\overline{\rm AH}$ is
harmonic and since its total flux through infinity, suitably
interpreted, equals the electric charge, we think of $-\Omega_{S^2}$
as representing the electric field of the proton. The comments made 
after \eqref{TNelectricfield} about the possibility of recovering a 
conventional electric field by contracting the electric 2-form on TN
with the vector field along asymptotic fibres apply, suitably
modified, to $\overline{\rm AH}$.

\begin{figure}[!ht]
\centering
\includegraphics[width=7truecm]{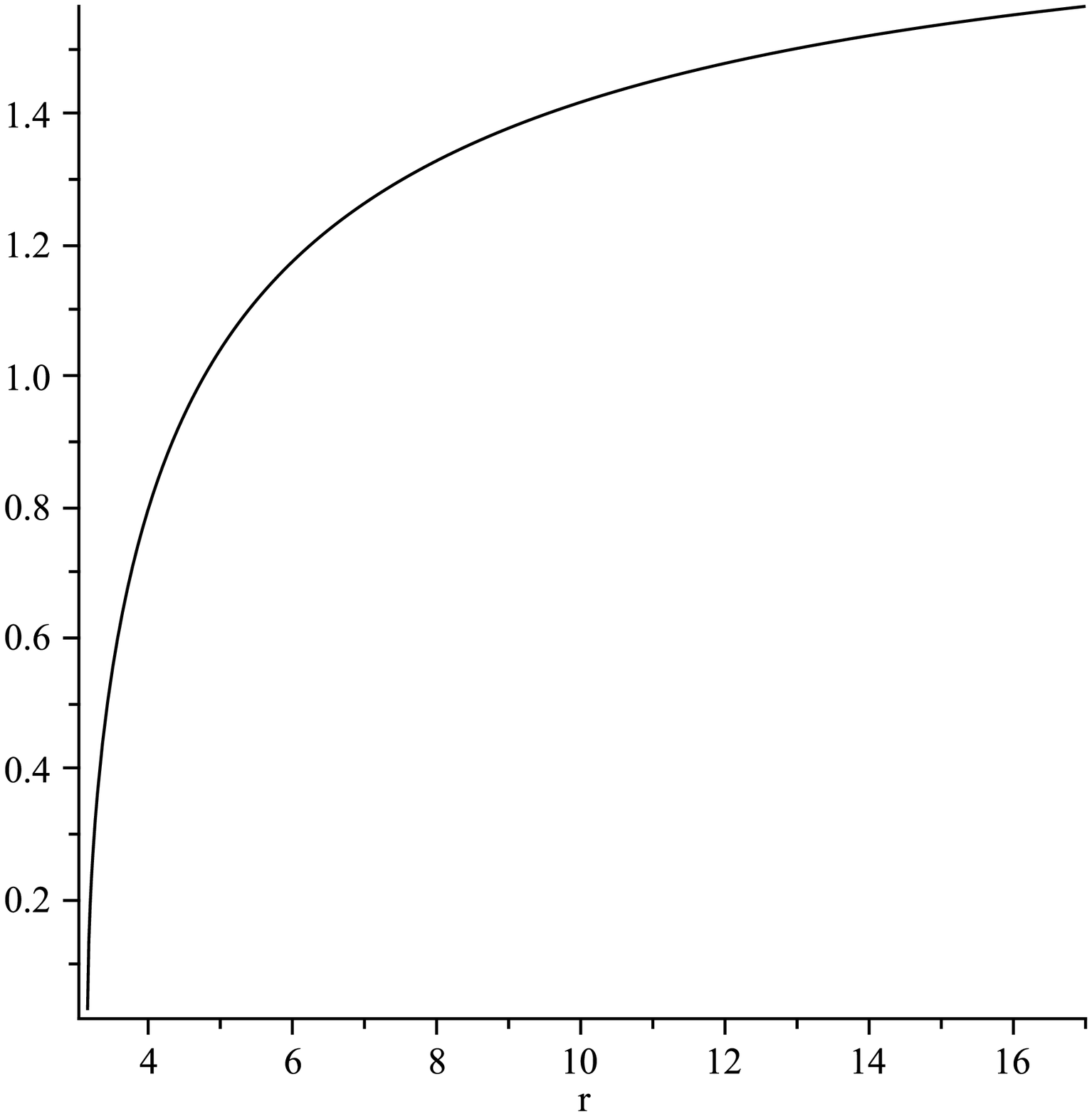}
\includegraphics[width=7truecm]{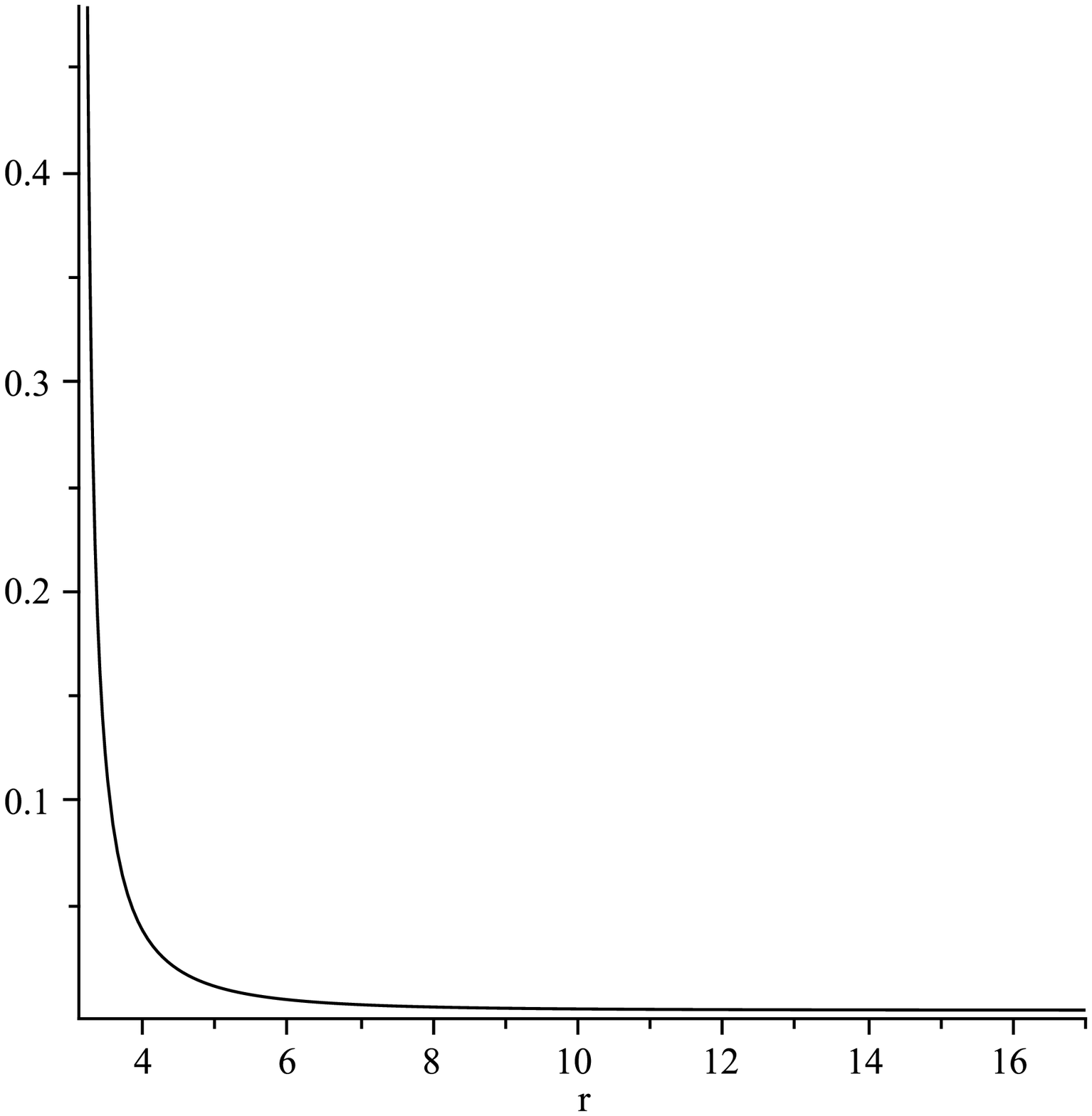}
\caption{The `electric' profile function $G_3^-$ and the energy
density $\rho_3^-$}
\label{AHelfield}
\end{figure}

In Fig.~\ref{AHelfield} we plot the numerically computed profile
function $G_3^-$ for $C=1$  (as defined in \eqref{smallg3}) and the
associated energy density \eqref{genen}, which turns out to be
\begin{equation}
\rho_3^-= 2\frac{(G_3^-)^2}{a^2b^2} \,.
\end{equation}
This function has a $(r-\pi)^{-1}$ singularity at $r=\pi$, and
falls off like $r^{-4}$ for large $r$ as in the TN case.

\section{Conclusion and outlook}

In introducing and illustrating geometric models of matter in
this paper we have concentrated on general, global properties of
particles like baryon number and electric charge. The most striking
phenomenological success of our geometric approach is its
prediction of precisely two stable electrically charged particles,
one leptonic and one baryonic, with opposite electric charges.
 
An important theme throughout this paper is the implementation of
rotational symmetry in our models. The rotation action  preserves the
metric and, in the compact cases, the distinguished surface where the
4-manifold intersects physical 3-space.
We have shown that our models are fermionic in the sense that the lift
of the rotation action to  spin bundles (over the `inside' of the
4-manifold in the compact cases) is necessarily an $SU(2)$-action, but
we have left the explicit construction of spin 1/2 states for future
work.
 
Non-vanishing electric charge and baryon number give rise to harmonic
2-forms on our model manifolds. These have allowed us to make
contact with the conventional description of charged particles in
terms of associated fields and fluxes.

There are various geometrically natural candidates for measuring
energy or mass in our models, but we have not committed ourselves to
any particular energy functional in this paper. In the absence of an
energy measure, we are not able to make contact with experimental data
about particle masses or forces between particles.
 
We end our summary with some observations about relative scales
predicted by our models for charged particles.
The AH model relates three scale parameters: the size of the core
(radius $\approx \pi$ in our geometric units), the size of the
asymptotic circles (radius 2 in geometric units) and the scale $1$
implicit in the exponential corrections to the asymptotic form of the
coefficient functions  \eqref{AHcoef} for the AH metric, which are
proportional to $e^{-r}$ (see \cite{GibbonsManton,Schroers} for a
discussion of these corrections in the context of magnetic
monopoles). Interpreting the core radius as the proton radius and
the scale in the exponential decay as the Compton wavelength
$\lambda_\pi$ of the pion, we find that our model correctly assigns
the same order of magnitude to those two quantities. The details are
not quite right (experimentally, the proton radius is just over half
of $\lambda_\pi$) but this is not surprising given our very rudimentary
understanding of how our model relates to experiment. In any case,
these considerations suggest that we should  pick a length unit  of
about 1 fermi or $10^{-15}$ m in the AH model.
 
Since the asymptotic fibration by circles arises for all electrically
charged particles, we expect the size of the asymptotic circle to have
a purely electric interpretation, and we also assume that it is the
same for both AH and TN (this was implicit in the way we fixed scale
and units for TN). One natural guess, at least for TN, is to relate
the size of the asymptotic circle to the classical electron
radius. Our models would then equate the orders of magnitude of the
classical electron radius with the Compton wavelength of the pion,
which is phenomenologically right. It is an attractive feature of TN
as a model for the electron that it has a length scale, but no core structure.
 
Many avenues remain to be explored. In our geometric approach, fusion
and fission of nuclei as well as decay processes involving both baryons
and leptons (like the beta decay of the neutron) should have a
description in terms of gluing and deformation of self-dual
4-manifolds. In order to study masses and binding energies of
particles we need to pick an energy measure. This could involve the
norms of curvatures and harmonic forms discussed here, but may take
a less conventional form. Our ideas about spin 1/2 need to be fleshed
out. The Dirac operator on our model manifolds is likely to play a
role in combining energy and spin considerations, and Seiberg-Witten
theory on the model manifolds seems relevant, too. In order to move
beyond static consideration, time needs to be introduced.
 
The list of open issues may seem daunting, but in each case the
geometric framework introduced here suggests natural lines of
attack. We hope to pursue them in future work.
 
\vspace{0.5cm}

\noindent {\bf Acknowledgment} \; We  thank Nigel Hitchin and 
Claude LeBrun for assistance and Jos\'e Figueroa-O'Farrill for
working with us during the early stages of the project. 

\vspace{0.5cm}
\appendix
\vspace{0.5cm}
\leftline{\bf\large {Appendices}}

\section{Geometry and sign conventions}
\label{conventions}
\subsection{Self-intersection numbers}
  
Let $L$ be a complex line bundle over a compact oriented surface $X$. Then
\bee
\label{selfintersect}
c_1(L) [X] = X^2            
\eee                             
where $X^2$ denotes the self-intersection number of the zero section
in the total space of $L$. This is essentially one of the definitions of
the first Chern class. Note that both $X$ and
$L$ are oriented, with the fibres having the natural orientation of
the complex numbers. However the self-intersection $X^2$ depends only
on the orientation of the total space, since $(-X)^2 = X^2$.

\subsection{The Hopf bundle}

Let us review the definition of the Hopf bundle $H$, the standard line
bundle over the complex projective line, paying particular attention
to orientations. We start with $\CC^2$, and $\CP^1$ as the space of
complex lines through the origin. This complex line bundle is defined
to be the dual of the line bundle $H$. The reason for this apparently
perverse choice is that $H$ has holomorphic sections (given by $\CC^2$)
and its dual $H^*$ does not. This is clear from the exact sequence of
holomorphic vector bundles over $\CP^1$:
\bee
0 \rightarrow H^* \rightarrow \CC^2  \rightarrow H  \rightarrow 0 \,.
\eee
Since holomorphic intersections are always positive, the first Chern
class of $H$ gives $+1$ when evaluated on the fundamental class of
$\CP^1$.

If we compactify $\CC^2$ by adding a $\CP^1$ at infinity then the
self-intersection number of this line is clearly $+1$, so that the
normal bundle is $H$.

\subsection{Non-orientable surfaces}

Now assume that $X$ is a non-orientable surface in an oriented
4-manifold $M$. This defines a mod 2 homology class in $M$ and so it
would appear that its self-intersection number is only defined modulo
2. However a more careful look shows that we can define an integer
self-intersection. There are several (equivalent) ways to see
this. First we note that the tangent bundle and the normal bundle of
$X$ in $M$ have isomorphic orientation real line bundles $L$. Then
both the Euler class of the normal bundle and the fundamental class of
$X$ are defined as classes `twisted' by $L$. The evaluation
\eqref{selfintersect} therefore makes sense and defines $X^2$. An
alternative way is to pass to the double covering of a neighbourhood
of $X$ in $M$, where $X$ acquires an orientation, compute the
self-intersection there and divide by 2. A third way is to deform $X$
into a transverse surface $\widetilde{X}$ and, near each intersection point of
$X$ and $\widetilde{X}$, choose an orientation of $X$ and the corresponding
orientation of $\widetilde{X}$ given by the deformation. Now compute the local
intersection numbers of $X$ and $\widetilde{X}$ and sum them up.
  
\subsection{An example}
\label{branched1}

The example we want to study is that of the real projective plane $X =
\RP^2$ inside the complex projective plane $M = \CP^2$. We give $M$ its
standard complex orientation. We plan to show that $X^2 = -1$.

To do this we will use the double branched covering of $\CP^2$ given
by the product of two copies of the complex projective line
$\CP^1$. We can identify $\CP^2$ with the symmetric product of the two
copies, branched along the diagonal $D$. This map is holomorphic and
compatible with the action of $SL(2,\CC)$. Now introduce a metric on
$\CP^1$ identifying it with a 2-sphere, and consider in 
$\CP^1 \times \CP^1$ the graph $D'$ of the antipodal map. The image 
of $D'$ in $\CP^2$ is the standard embedding of
$\RP^2$ into $\CP^2$. Outside the branch locus $D$ the map from the
product of the two 2-spheres to $\CP^2$ is a double covering so we can
compute $X^2$ as half of $(D')^2$. This is most easily done by
observing that, in the cohomology of the product, with $x$ and $y$ the
two generators of $H^2$, the classes of $D$ and $D'$ are
\bee
D = x + y \,, \quad  D' = x - y \,.
\eee
Since $x^2 = y^2 = 0$ and $xy =1$  it follows that  
\bee
D^2 = 2 \quad \text{and} \quad   (D')^2  = -2
\eee
(while $D.D' = 0$ agreeing with the fact that $D$ and $D'$ are disjoint).
Dividing by 2 to get back down to $\CP^2$ we end up with  $X^2 = -1$ as stated.

As a check we can also use a deformation of $D'$ in $M $ by choosing
a different metric on $\CP^1$ giving a different antipodal map. In particular,
consider the 2-sphere as the boundary of the unit ball in $\RR^3$ and
move the origin so that antipodal points are now the end points of
lines through the new origin. The line joining the two origins
defines a unique intersection point of the two copies of $\RP^2$ in
$\CP^2$. Since this line acquires different orientations from the two
origins the intersection number is $-1$.
 
A third sign check is to observe that the normal bundle to the
diagonal in the product of the two 2-spheres is the tangent bundle
with Euler number 2, while for the antidiagonal (graph of the
antipodal map) it is the cotangent bundle with Euler number $-2$.

\subsection{More on the example}
\label{branched2}

It is enlightening to analyze the $SO(3)$ action on the spaces in the
example above. In addition we can introduce the branched double covering
$\CP^2 \rightarrow S^4$ given by complex conjugation
\cite{Massey,Kuiper}. This is branched over $\RP^2$.
 
Thus we have maps
\bee
S^2 \times S^2 \rightarrow \CP^2 \rightarrow S^4 \,,
\eee
which are compatible with the $SO(3)$ actions. Here $S^4$ is best
thought of as the space of real $3\times 3$ symmetric matrices of
zero trace and fixed norm.

The generic orbits are, in each case, 3-dimensional but there are 
two exceptional orbits of dimension 2. Inside the product 
$S^2 \times S^2$ they are the diagonal and anti-diagonal (both $S^2$), 
in $S^4$ they are both $\RP^2$, while in $\CP^2$ one is $S^2$ and 
one is $\RP^2$. Each of the two maps is branched along a surface: 
$S^2$ for the first map and $\RP^2$ for the second.

To examine the self-intersection numbers of the exceptional orbits it
is useful to note two general rules:
\begin{enumerate}
\item[A.]  If $X$ in $M$ is double covered by $X'$ in $M'$  then
$(X')^2  = 2 X^2$.
\item[B.] If $X$ is a branch locus in the double covering then  $(X')^2
=\frac 1 2  X^2$.
\end{enumerate}
Rule A is obvious and rule B is familiar in algebraic geometry, but it
applies more generally even in the non-orientable case.

Using rules A and B we can now check what happens in our two maps. In $S^4$
the two $\RP^2$s have self-intersection numbers $+2$ and $-2$. In $S^2
\times S^2 $ the two 2-spheres also have self-intersection numbers
$+2$ and $-2$. In $\CP^2$ the 2-sphere, which is a  conic,
has self-intersection number $+4$, while $\RP^2$ has self-intersection
number $-1$.
  
\subsection{Self-dual manifolds}
\label{signsTNAH}

We want to analyze various sign conventions in relation to self-dual
manifolds, and begin by checking our sign conventions for two examples 
of complex K\"ahler surfaces. Referring to our discussion of the 
signature and Euler characteristic in Sect.~\ref{sigsect}, we note the examples:
\begin{enumerate}
\item The K3 surface with the Yau metric, which makes it
hyperk\"ahler. With its complex orientation its signature is $-16$
and its Euler characteristic is 24. With the opposite orientation, the metric is self-dual and  the
signature is positive.
\item $\CP^2$ with the Fubini-Study metric is self-dual for its
complex orientation \cite{EGH}, agreeing with the positive signature $+1$.
\end{enumerate}

We will also be interested in non-compact examples similar to the
above. The argument that a hyperk\"ahler manifold is self-dual for the
orientation opposite to the complex orientation (given by one of the
family of complex K\"ahler structures) is purely local. It just
depends on the fact that the bundle of self-dual 2-forms for these
complex orientations is flat. More generally, on any K\"ahler
manifold, the bundle of self-dual 2-forms is the direct sum of the
canonical line bundle $K$, its dual and a trivial bundle generated by
the K\"ahler form.

The first non-compact example that interests us is the Taub-NUT manifold  TN.
This has the isometry group $U(2)$, its topology (and symmetry) is
that of $\CC^2$ with the central $U(1)$ giving the scalar action. The
other $U(1)$ actions, inside $SU(2)$, define the complex hyperk\"ahler
structures which have the opposite orientation.  Hence TN is self-dual
for the orientation that becomes that of $\CC^2$ in the limit when the
parameter $\epsilon$ goes to 0, as discussed after \eqref{TNcoef}.
 
The second example is that of the (simply-connected version of the)
Atiyah-Hitchin manifold AH, which is an open subspace of $\CP^2$ got
by removing $\RP^2$. As shown in \cite{AtiyahHitchin}, it has a complete
hyperk\"ahler metric. This is self-dual for the orientation opposite
to the complex orientations given by the hyperk\"ahler metric. But
this is just the orientation given by its complex structure as an 
open subset of $\CP^2$. This can be checked directly, but it is best 
seen by using the results of Hitchin \cite{Hitchin}, reviewed in Sect.~\ref{Hitchinsection}, 
which give  a whole sequence of self-dual manifolds on the same space, 
starting from the Fubini-Study metric of $\CP^2$ and converging to AH.

In each case, for TN and AH, we have a hyperk\"ahler manifold acted on
by $SU(2)$ (the action descends to $SO(3)$ for AH). The manifolds have
a `core region' and an `asymptotic region' with an asymptotic action of
$O(2)$. In addition to the 2-parameter family of complex structures
given by the hyperk\"ahler metric (and rotated by $SU(2)$) there is
another complex structure compatible with the $SU(2)$ action, but
giving the opposite orientation to that of the 2-parameter family.
For TN this complex manifold is just $\CC^2$ and for AH it is $\CP^2
\setminus \RP^2$.

In both cases the asymptotic $O(2)$ action gives us topologically a
disc bundle at infinity  and this can be identified with the normal
bundle of the surface $X$  that naturally compactifies the manifold.
For TN this compactification is $\CP^2$ with $X = \CP^1$, and for 
AH it is $\CP^2$ with $X=\RP^2$.  We have shown that these give rise to 
opposite signs:
\[  
\mbox{For TN} \quad X^2 = +1 \quad  \mbox{while for AH} \quad X^2 = -1 \,.
\] 
As explained in the main text,  TN  and AH are our  geometric models for  the electron and  the proton, and we  identify the self-intersection number of $X$ with minus the electric charge. 

\section{Metric properties of TN and AH}
\label{AHTNdetails}
\subsection{Coordinates and conventions}

In this paper we need to parametrise $SU(2)$ explicitly at various
points, and we use generators
\bee
t_1=\frac{1}{2}i\tau_1 = \frac{1}{2}
\begin{pmatrix} 0 &i  \\ i  &  0  \end{pmatrix} \,, \quad 
t_2=\frac{1}{2}i\tau_2 = \frac{1}{2}
\begin{pmatrix}  \phantom{-}0 &1  \\ -1  &  0 \end{pmatrix} \,, \quad 
t_3=\frac{1}{2}i\tau_3 = \frac{1}{2}
\begin{pmatrix} i & \phantom{-}0  \\ 0 & -i \end{pmatrix}
\,,
\eee
where $\tau_a$, $a=1,2,3$, are the Pauli matrices; the commutators are
$[t_a,t_b]=-\epsilon_{abc}t_c$. We use Euler angles 
$\theta\in [0,\pi)$, $\phi\in[0,2\pi)$ and $\psi\in[0,4\pi)$
for parametrising $SU(2)$ as follows:
\bee
g(\phi,\theta,\psi)= e^{\phi t_3}e^{\theta t_2 } e^{\psi t_3}= \begin{pmatrix}
e^{\frac i 2 (\psi + \phi)} \cos \frac{1}{2}\theta & e^{\frac i 2 (\phi-\psi)} 
\sin \frac{1}{2}\theta \\
-e^{\frac i 2 (\psi-\phi)} \sin \frac{1}{2}\theta &
e^{-\frac i 2 (\psi + \phi)} \cos \frac{1}{2}\theta \end{pmatrix}.
\eee
Defining left-invariant 1-forms $\eta_a$ on $SU(2)$ via
\begin{equation}
\label{leftinvariant}
g^{-1}\d g = \eta_1 t_1 + \eta_2 t_2 + \eta _3 t_3 \,,
\end{equation}
we compute to find
\begin{align}
\label{linv}
\eta_1 & \ = \ - \sin \psi \, \d \theta + \cos \psi \sin \theta \, \d \phi \,, 
\nonumber \\
\eta_2 & \ = \ \phantom{-} \cos \psi \, \d \theta 
+ \sin \psi \sin \theta \, \d \phi \,,  \nonumber \\
\eta_3 & \ = \ \phantom{-} \, \d \psi + \cos \theta \, \d \phi \,, 
\end{align}
satisfying $\d \eta_i = \half \epsilon_{ijk} \eta_j \wedge \eta_k$. 
They descend to $SO(3)$ by simply restricting $\psi$ to $[0,2\pi)$. 

Explicitly, generators of the Lie algebra of $SO(3)$ are
$(T_a)_{bc}=\epsilon_{abc}$. They also satisfy
$[T_a,T_b]=-\epsilon_{abc}T_c$. We can then parametrise $SO(3)$
matrices via
\bee
\label{rotpara}
R(\phi,\theta,\psi)= e^{\phi T_3}e^{\theta T_2 } e^{\psi T_3},
\eee
with $\theta\in [0,\pi)$, $\phi\in[0,2\pi)$ and $\psi\in[0,2\pi)$.
Then 
\begin{equation}
\label{Leftinvariant}
R^{-1}\d R = \eta_1 T_1 + \eta_2 T_2 + \eta _3 T_3
\end{equation}
provides an alternative definition of the left-invariant 1-forms 
on $SO(3)$: one obtains the same expressions as in \eqref{linv}, but 
with the range of angles automatically appropriate to $SO(3)$.

Both TN and AH can be parametrised in terms of Euler angles and a
radial coordinate. For TN, the angular ranges are $\theta \in [0,\pi)$, 
$\phi\in [0,2\pi)$ and $\psi \in [0,4\pi)$. For AH they are 
$\theta \in [0,\pi)$, $\phi\in [0,2\pi)$ and $\psi \in [0,2\pi)$
with the additional $\ZZ_2$ identification
\bee
\label{ident}
(\theta,\phi,\psi )\simeq (\pi- \theta,\phi+\pi,-\psi),
\eee
which, in the asymptotic region, is the simultaneous reversal of
spatial and fibre direction. The following angular integrals, which 
enter into various calculations in this paper, are therefore
\begin{align}
\label{angularintegrals}
\int_{SU(2)}\eta_1\wedge\eta_2\wedge\eta_3 &= -16\pi^2 \,, \nonumber \\
\int_{SO(3)}\eta_1\wedge\eta_2\wedge\eta_3 &= -8\pi^2 \,, \nonumber \\
 \int_{(SO(3)/\ZZ_2)}\eta_1\wedge\eta_2\wedge\eta_3 &= -4\pi^2 \,. 
\end{align}
Note that $\eta_1$ is invariant under \eqref{ident}, but $\eta_2$ and
$\eta_3$ change sign.

Recalling that the metrics on TN and AH are of the Bianchi IX form 
\begin{align}
\label{BianchiIX}
ds^2 = f(r)^2dr^2 + a(r)^2\eta_1^2 +b(r)^2\eta_2^2 +c(r)^2\eta_3^2 \,, 
\end{align}
we introduce the tetrad
\bee
\label{tetrad}
\theta^1=a \eta_1 \,, \; \theta^2=b\eta_2 \,, \; \theta^3 = c\eta_3 \,, \;  
\theta^4=f dr \,. 
\eee
The self-duality of the Riemann tensor computed from
\eqref{BianchiIX} with respect to the volume element
\bee
\label{volumedef}
\d V = a\eta_1\wedge b\eta_2\wedge c\eta_3\wedge f\d r =
-fabc \, dr \wedge\eta_1 \wedge\eta_2 \wedge\eta_3
\eee
is then equivalent to the set of ordinary differential equations 
\bee
\label{gendual}
\frac {2bc}{f}\frac { da}{dr} = (b-c)^2 -a^2 +2\lambda bc \,, \qquad 
\mbox{+ cyclic} \,,
\eee
where `+ cyclic' means we add the two further equations
obtained by cyclic permutation of $a,b,c$, and $\lambda$ is a
parameter which has to be either $0$ or $1$. In all cases the
resulting metrics are hyperk\"ahler, but in order to obtain metrics
whose hyperk\"ahler structures are rotated by the $SU(2)$ action we
need to set $\lambda=0$. This is the case for both TN and AH, so $\lambda=0$
in \eqref{gendual}.
 
One checks the equivalence of the self-duality of the Riemann tensor  
and \eqref{gendual} as follows. Denoting the Riemann tensor by $R$, 
its component 2-forms with respect to \eqref{tetrad} are 
\begin{align}
R_{12}&=\d \mu_3 \wedge \eta_3 + (\mu_3-\mu_1\mu_2-\lambda_1\lambda_2)
\eta_1\wedge\eta_2 \,, \nonumber  \\
R_{34}&=\d \lambda_3 \wedge \eta_3
+(\lambda_3-\lambda_1\mu_2-\lambda_2\mu_1)\eta_1\wedge\eta_2 \,,
\end{align}
and similar expressions for the components obtained by cyclic
permutation of $1,2,3$, where
\bee
\label{mfcts}
\mu_1= \frac{b^2+c^2 -a^2}{2bc} \,,\quad \mu_2= 
\frac{c^2+a^2 -b^2}{2ca} \,,\quad \mu_3= \frac{a^2+b^2 -c^2}{2ab}
\eee
and 
\bee
\lambda_1=\frac{1}{f}\frac{da}{dr} \,, \quad  
\lambda_2=\frac{1}{f}\frac{db}{dr} \,, \quad
\lambda_3=\frac{1}{f}\frac{dc}{dr} \,.
\eee
When \eqref{gendual} hold, then $\lambda_i=\mu_i+\lambda -1$, $i=1,2,3$,
with $\lambda=0$ or $\lambda =1$, and the self-duality relations
\bee
R_{12}=R_{34} \,, \quad R_{23}=R_{14} \,, \quad R_{31}=R_{24}
\eee
are easily verified.
One also checks that each of the three independent 2-forms $R_{12},R_{23}, 
R_{31}$ is self-dual with respect to \eqref{volumedef}, as expected. 

The only solutions of \eqref{gendual} with $\lambda=0$ which give rise 
to complete manifolds whose generic $SU(2)$- or $SO(3)$-orbit is 
three-dimensional are the TN
and AH metrics, whose coefficient functions we discussed in
Sect.~\ref{protonelectron}. It is important that the coefficient 
function $c$ in \eqref{athi} is negative for all $r$ in the AH
manifold. It implies in particular that the canonical volume 
element/orientation $\d V$ in \eqref{volumedef}, with $f = -b/r$,
\bee
\label{volume}
\d V = -fabc \, \d r\wedge  \eta_1\wedge  \eta_2\wedge \eta_3 
= -\frac{ab^2c }{r} \sin \theta \, \d r\wedge  \d \theta\wedge 
\d \phi \wedge \d \psi \,,
\eee
has opposite signs for TN and AH, in the following sense.  
Assuming the same orientation $\eta_1 \wedge \eta_2 \wedge \eta_3$ of
the $SU(2)$ and  $SO(3)$ orbits in TN and, respectively, AH, the radial 
direction is oppositely oriented in the two cases: the natural radial line
element $-fabc \, \d r$ has the same orientation as $\d r$ for TN, but the
opposite orientation for AH. This will be important when evaluating
various integrals. It means that, when using the coordinates
$r,\theta,\phi,\psi$ we can use the conventional orientations for
these coordinates when computing for AH, but should use the opposite
orientation when computing for TN. Thus we integrate in the negative
$r$-direction when calculating on TN.
 
\subsection{$L^2$-norms of the Riemann curvature}
\label{Riemanncalculations}

The squared $L^2$-norm  of the Riemann tensor is
\bee
\label{curvsquared}
||R||_M^2 = \int_{M} 2(R_{12}\wedge R_{12} + R_{23}\wedge R_{23}+
R_{31}\wedge R_{31}) \,,
\eee
where $M$ is TN or AH, and we have used the self-duality of the curvature.
In terms of the functions \eqref{mfcts} the integrand can be
simplified (see also \cite{SethiSternZaslow}, where this calculation
is carried out for the AH metric) and becomes
\begin{equation}
\label{integral}
2(R_{12}\wedge R_{12} + R_{23}\wedge R_{23}+ R_{31}\wedge R_{31})=
dF\wedge\eta_1\wedge\eta_2\wedge\eta_3,
\end{equation}
with
\bee
F=2(\mu_1+\mu_2+\mu_3-1)^2-8\mu_1\mu_2\mu_2 \,. 
\eee
We denote these functions in the TN and AH cases by $F_{{\rm TN}}$ and
$F_{{\rm AH}}$. 

For TN, we can compute explicitly and find 
\bee
F_{{\rm TN}}(r)= -8 \, \frac{2r+1}{(r+2)^4} \,.
\eee
Integrating \eqref{integral} and using \eqref{angularintegrals} as
well as our sign convention for the volume form we find
\bee
||R||^2_{{\rm TN}} =(-16\pi^2) (F_{{\rm TN}}(0)-F_{{\rm TN}}(\infty))
= 8\pi^2 \,.
\eee

For AH, similarly,
\bee
||R||^2_{{\rm AH}} =(-4\pi^2) (F_{{\rm AH}}(\infty)-F_{{\rm AH}}(\pi))
\,.
\eee
To complete the calculation we need the behaviour of $F_{{\rm AH}}$ at infinity
and at $\pi$. One finds for AH
\begin{align}
\mu_1(\infty)=\mu_2(\infty) =\frac{c}{2a}\left(\infty\right)= 0 \,,\quad
\mu_3(\infty)= 1-\frac{c^2}{2a^2}\left(\infty\right) = 1 \,,
\end{align}
as well as 
\bee
\mu_1(\pi)=-1 \,, \quad \mu_2(\pi) =\frac{1}{2} \,,\quad \mu_3(\pi)= 
\frac 1 2 \,,
\eee
where we took careful limits, using \eqref{AHcoef} and \eqref{AHsmallr}.
Therefore 
\begin{equation}
\label{AHboundaries}
F_{{\rm AH}}(\infty)=0 \,, \qquad F_{{\rm AH}}(\pi)= 4 \,,
\end{equation}
so
\bee
||R||^2_{{\rm AH}} = (-4\pi^2)\times(0-4) =16 \pi^2 \,.
\eee

\section{Harmonic forms on TN and AH}
\label{harmonicsect}
\subsection{Rotationally symmetric harmonic forms}

The question of computing harmonic 2-forms on the TN and AH manifolds
arose in physics in the context of testing S-duality, see \cite{Sen}
for   AH and \cite{LeeWeinbergYi,GauntlettLowe} for TN. Harmonic
forms also play a role as curvatures of line bundles over TN and AH,
see \cite{MantonSchroers} for a discussion of this for AH where the
form later used for testing S-duality \cite{Sen}  is the curvature of an index bundle.
This appendix contains a systematic discussion of these forms and also
a harmonic form on the branched cover of AH which has not previously
been considered in the literature. The physical interpretation of the
forms as electric and baryonic fluxes is given in Sect.~\ref{enfluxsection}.

On both TN and AH, rotationally symmetric 2-forms can be written in terms 
of the metric coefficient functions $f,a,b,c$ appearing in the 
respective metrics and functions $f_1,f_2$ and $f_3$ of the radial 
coordinate $r$ as
\begin{align}
\label{ASDforms}
\omega^{\pm}_1&=f^\pm_1 \left( b \eta_2\wedge c\eta_3  \pm  a \eta_1
\wedge  f\d r \right) \,, \nonumber \\
\omega^{\pm}_2&=f^\pm_2 \left(  c \eta_3\wedge a\eta_1   \pm b \eta_2
\wedge  f \d r \right) \,, \nonumber \\
\omega^{\pm}_3&=f^\pm_3 \left( a \eta_1\wedge b\eta_2 \pm c \eta_3 
\wedge  f \d r\right) \,,
\end{align}
with $+$ standing for self-dual and $-$ standing for anti-self-dual.
Introducing the functions
\bee
\label{coeffren}
g_1^\pm = f_1^\pm bc \,, \quad g_2^\pm = f_2^\pm ca \,, \quad 
g_3^\pm = f_3^\pm ab \,,
\eee
closure of the forms \eqref{ASDforms} implies the ordinary
differential equations
\begin{align}
\label{closed}
\frac{\d g_1^\pm}{\d r} &= \mp \frac{af}{bc} g_1^\pm \,,\nonumber \\
\frac{\d g_2^\pm}{\d r} &= \mp \frac{bf}{ca} g_2^\pm \,, \nonumber \\
\frac{\d g_3^\pm}{\d r} &= \mp \frac{cf}{ab} g_3^\pm \,.
\end{align}
For solutions of these differential equations (we shall discuss below
in which cases regular solutions exist), the forms \eqref{ASDforms}
can be written
\begin{align}
\label{formsagain}
\omega^{\pm}_1&=g^\pm_1\eta_2\wedge \eta_3 + \d g^\pm_1\wedge \eta_1 \,,
\nonumber \\
\omega^{\pm}_2&=g^\pm_2  \eta_3\wedge \eta_1 + \d  g^\pm_2\wedge
\eta_2 \,, \nonumber \\
\omega^{\pm}_3&=g^{\pm}_3 \eta_1\wedge \eta_2 + \d g^{\pm}_3\wedge
\eta_3 \,. 
\end{align}
All these forms are locally exact, that is,
\bee
\omega^{\pm}_i=\d(g^\pm_i  \eta_i) \,, \quad i=1,2,3 \,,
\eee
but the 1-forms in brackets may not be globally defined and, when 
defined, they may not be $L^2$, so that the corresponding forms
$\omega^\pm_i$ are not necessarily  $L^2$-exact.

We are interested in regular and bounded solutions of the equations
\eqref{closed}. Using our convention $f=-b/r$, the functions appearing
in the differential equations are then
\bee
\label{coeffun}
\frac{af}{bc} = -\frac{a}{rc}\,, \quad  \frac{bf}{ca} =
-\frac{b^2}{rca}\,, \quad \frac{cf}{ab} = -\frac{c}{ra}\,.
\eee

\subsection{Taub-NUT}

Substituting the TN coefficient functions \eqref{TNcoef} (with
$\epsilon=1$ and $m=2$), the functions \eqref{coeffun} simplify further to 
\bee
-\frac{a}{rc}=-\frac{b^2}{rca}=-\frac{r+2}{2r} \,, 
\quad -\frac{c}{ra}= -\frac{2}{r(r+2)} \,.
\eee
It is easy to check, and was shown in \cite{LeeWeinbergYi} and 
\cite{GauntlettLowe}, and earlier in \cite{Brill} and \cite{Pope},
that only the equation  for $g^+_3$  has a solution which is both 
regular at the origin and bounded as $r\rightarrow \infty$. 
Integrating, one finds explicitly
\bee
g^+_3(r) =C\frac{r}{r+2} \,,
\eee
with an arbitrary constant $C$. Setting $C=1$, \eqref{formsagain} gives 
the associated normalisable, harmonic form
\bee
\omega_3^+ =\left( \frac{r}{r+2} \, \eta_1\wedge \eta_2  
+ \frac{2}{(r+2)^2} \, dr \wedge \eta_3\right) 
=  d\left(\frac {r}{r+2} \, \eta_3\right) \,.
\eee
Then, using \eqref{angularintegrals}, 
\bee
\int_{{\rm TN}} \omega_3^+\wedge\omega_3^+ = \int_\infty^0
\frac{4r}{(r+2)^3} \, \d r\times  (-16 \pi^2) = 16 \pi^2 \,.
\eee

\subsection{Atiyah-Hitchin}

On AH, the analysis of harmonic forms is analogous. We write $G^\pm_i$
for the solutions of \eqref{closed} with the radial functions
$a,b,c,f$ of the AH metric, and write the forms obtained by solving
these equations as
\begin{align}
\label{formsagainAH}
\Omega^{\pm}_1&=G^\pm_1\eta_2\wedge \eta_3  + \d G^\pm_1\wedge \eta_1 \,, 
\nonumber \\
\Omega^{\pm}_2&=G^\pm_2\eta_3\wedge \eta_1 + \d  G^\pm_2 \wedge \eta_2 \,, 
\nonumber \\
\Omega^{\pm}_3&=G^{\pm}_3 \eta_1\wedge \eta_2 + \d G^{\pm}_3
\wedge\eta_3 \,. 
\end{align}
As for TN,  all these forms are  formally exact, 
\bee
\Omega^{\pm}_i=\d(G^\pm_i  \eta_i),\quad i=1,2,3 \,,
\eee
but the 1-forms in brackets may not be globally defined, as we 
shall see.
 
Note that only the forms $\Omega^\pm_1$ respect the identification
\eqref{ident}. However, as explained in the main text we also consider
the  branched double cover $\overline{\rm AH}$ of AH and therefore 
keep all forms in the
discussion. On AH, we use \eqref{AHsmallr} and \eqref{AHcoef} to find 
that the functions appearing in the differential equations 
\eqref{closed} have the following behaviour near the core 
(i.e. $(r-\pi)$ small),
\bee
\label{AHbolt}
\frac{a}{rc} \sim -\frac{2}{\pi^2}(r-\pi) \,, \quad 
\frac{b^2}{rca} \sim \frac{c}{ra} \sim -\frac{1}{2(r-\pi)} \,,
\eee
and the following behaviour for large $r$,
\bee
\label{AHinfty}
\frac{a}{rc}
\sim \frac{b^2}{rca} \sim -\frac{1}{2}\left(1- \frac 2 r\right), \quad
\frac{c}{ra} \sim -\frac{2}{r(r-2)} \,.
\eee
For the differential equations \eqref{closed} on AH, this means that only
the equations for $G^+_1$ and  $G^-_3$ have solutions which are finite
at the core and remain bounded for large $r$. Both of these solutions are
discussed and interpreted in the main text.
 
For any solution of \eqref{closed} on AH, we note 
\begin{align}
\label{masterintAH}
\int_{{\rm AH}}\Omega^\pm \wedge\Omega^\pm &= \int_{{\rm AH}}\d((G^\pm)^2)
\wedge \eta_1 \wedge\eta_2\wedge\eta_3 \nonumber \\ 
&= (-4\pi^2)\left((G^\pm)^2(\infty)-(G^\pm)^2(\pi)\right) \,,
\end{align}
where $\Omega^\pm$ and $G^\pm$ stand for any of the forms and
coefficient functions in \eqref{formsagainAH}, and
we again used \eqref{angularintegrals}. We also note that the
$L^2$-norm of each of these forms can be written as
\begin{align}
||\Omega^{\pm}||^2 &= \pm \int_{{\rm AH}}\Omega^\pm \wedge
\Omega^\pm  \nonumber \\
&= \int_{{\rm AH}} \rho \, \d V \,,
\end{align}
where $\d V$ is the volume element defined in \eqref{volumedef} and $\rho$ is
a density which we interpret as an energy density in the main
text. Its general form is
\begin{equation}
\label{genen}
\rho =  \mp \frac{2}{fabc} \, G^\pm \, \frac{\d G^\pm}{\d r} \,,
\end{equation}
which can be simplified in each of the cases, using the differential 
equations \eqref{closed} on AH.


\begin{thebibliography}{99}

\bibitem{Skyrme}
T.~H.~R.~Skyrme, A non-linear field theory,
Proc. Roy. Soc. A260 (1961) 127--138.

\bibitem{BrownRho}
G.~E.~Brown and M.~Rho, {\em {The Multifaceted Skyrmion}},
Singapore, World Scientific, 2010.

\bibitem{Penrose} 
R.~Penrose, The twistor programme, Rep. Math. Phys. 12 (1977) 65--76.

\bibitem{DonaldsonFriedman}
S.~Donaldson and R.~Friedman, Connected sums of self-dual manifolds
and deformations of singular spaces, Nonlinearity 2 (1989) 197--239.

\bibitem{AtiyahManton}
M.~F.~Atiyah and N.~S.~Manton, Skyrmions from instantons, 
Phys. Lett. B222 (1989) 438--442.

\bibitem{SakaiSugimoto}
T.~Sakai and S.~Sugimoto, Low energy hadron physics in holographic 
QCD, Prog. Theor. Phys. 113 (2005) 843--882.

\bibitem{Sutcliffe}
P.~Sutcliffe, Skyrmions, instantons and holography, JHEP 1008:019 (2010). 

\bibitem{AtiyahSutcliffe}
M.~Atiyah and P.~Sutcliffe, Skyrmions, instantons, mass and
curvature, Phys. Lett. B605 (2005) 106--114. 

\bibitem{AtiyahSinger}  M.~F.~Atiyah and I.~M.~Singer, The index of 
elliptic operators: III, Ann. of Math. 87 (1968) 546--604.

\bibitem{Minerbe}
V.~Minerbe, On the asymptotic geometry of gravitational instantons, 
Ann. Scient. \'Ec. Norm. Sup. (4) 43 (2010) 883--924. 

\bibitem{Hawking}
S.~W.~Hawking, Gravitational instantons,
Phys. Lett. A60 (1977) 81--83.

\bibitem{EGH} 
T.~Eguchi, P.~B.~Gilkey and A.~J.~Hanson, Gravitation, 
gauge theories and differential geometry, Physics Reports 66 (1980) 213--393.
  
\bibitem{AtiyahHitchin}
M.~Atiyah and N.~Hitchin, {\em {The Geometry and Dynamics of
Magnetic Monopoles, M.~B.~Porter Lectures, Rice University}}, 
Princeton NJ, Princeton University Press, 1988.
  
\bibitem{Pollard} 
D.~Pollard, Antigravity and classical solutions of 
five-dimensional Kaluza-Klein theory, J. Phys. A16 (1983) 565--574.

\bibitem{GrossPerry} 
D.~J.~Gross and M.~J.~Perry, Magnetic monopoles in 
Kaluza-Klein theories, Nucl. Phys. B226 (1983) 29--48.

\bibitem{Sorkin} 
R.~D.~Sorkin, Kaluza-Klein monopole, Phys. Rev. Lett. 51 (1983) 87--90.
  
\bibitem{GibbonsPope}
G.~W.~Gibbons and C.~N.~Pope, $\CP^2$ as a gravitational instanton, 
Commun. Math. Phys. 61 (1978) 239--248.

\bibitem{BouchiatGibbons}
C.~Bouchiat and G.~W.~Gibbons, Non-integrable quantum phase in the 
evolution of a spin-1 system: a physical consequence of the non-trivial
topology of the quantum state-space, J. Phys. France 49 (1988) 187--199.
  
\bibitem{DancerStrachan}
A.~S.~Dancer and I.~A.~B.~Strachan, K\"ahler-Einstein metrics with 
$SU(2)$ action, Math. Proc. Camb. Phil. Soc. 115 (1994) 513--525.
  
\bibitem{AtiyahMantonCMP} 
M.~F.~Atiyah and N.~S.~Manton, Geometry and kinematics of two 
Skyrmions, Commun. Math. Phys. 153 (1993) 391--422.

\bibitem{Hitchin}
N.~J.~Hitchin, A new family of Einstein metrics, in {\em {Manifolds and 
Geometry (Pisa, 1993), p. 190--222, Sympos. Math. XXXVI}}, 
Cambridge, Cambridge University Press, 1996.

\bibitem{Hitchin2} 
N.~J.~Hitchin, Twistor spaces, Einstein metrics and isomonodromic 
deformations, J. Diff. Geom. 42 (1995) 30--112 . 

\bibitem{AtiyahLeBrun}
M.~F.~Atiyah and C.~LeBrun, The signature of 4-manifolds with conical 
singular metric, in preparation. 

\bibitem{GibbonsRuback}
G.~W.~Gibbons and P.~J.~Ruback, The hidden symmetries of multi-center
metrics, Commun. Math. Phys. 115 (1988) 267--300.

\bibitem{Olivier}
D.~Olivier, Complex coordinates and K\"ahler potential for the
Atiyah-Hitchin metric, Gen. Rel. and Grav. 23 (1991) 1349--1362.

\bibitem{AtiyahPatodiSinger} 
M.~F.~Atiyah, V.~K.~Patodi and I.~M.~Singer, Spectral asymmetry and 
Riemannian geometry: I, Math. Proc. Camb. Phil. Soc. 77 (1975) 43--69.
  
\bibitem{LeBrun}
C.~LeBrun, Curvature functionals, optimal metrics, and the
differential topology of 4-manifolds, in {\em {Different Faces of Geometry}}, 
S.~Donaldson, Ya.~Eliashberg and M.~Gromov, eds., 
New York, Kluwer Academic/Plenum, 2004. 

\bibitem{Brill} 
D.~R.~Brill, Electromagnetic fields in a homogeneous, nonisotropic
universe, Phys. Rev. 133 (1964) B845--B848.
 
\bibitem{Pope} 
C.~N.~Pope, Axial-vector anomalies and the index theorem in charged
Schwarzschild and Taub-NUT spaces, Nucl. Phys. B141 (1978) 432--444. 
  
\bibitem{LeeWeinbergYi}
K.~Lee, E.~J.~Weinberg and P.~Yi, Electromagnetic duality and $SU(3)$
monopoles, Phys. Lett. B376 (1996) 97--102.  
  
\bibitem{GauntlettLowe}
J.~P.~Gauntlett and D.~A.~Lowe, Dyons and S-duality in $N=4$ supersymmetric
gauge theory, Nucl. Phys. B472 (1996) 194--206.

\bibitem{MantonSchroers}
B.~J.~Schroers and N.~S.~Manton, Bundles over moduli spaces and 
the quantisation of BPS monopoles, Annals of Physics 225 (1993) 290--338.

\bibitem{Sen}
A.~Sen, Dyon-monopole bound states, self-dual harmonic forms on the 
multi-monopole moduli space, and $SL(2,\ZZ)$ invariance in string 
theory, Phys. Lett. B329 (1994) 217--221.

\bibitem{GibbonsManton}
G.~W.~Gibbons and N.~S.~Manton, Classical and quantum dynamics of BPS
monopoles, Nucl. Phys. B274 (1986) 183--224. 

\bibitem{Schroers}
B.~J.~Schroers, Quantum scattering of BPS monopoles at low energy,
Nucl. Phys. B367 (1991) 177--214. 

\bibitem{Massey}
W.~S.~Massey, The quotient space of the complex projective plane under
conjugation is a 4-sphere, Geom. Dedicata 2 (1973) 371--374.

\bibitem{Kuiper}
N.~H.~Kuiper, The quotient space of $\CP(2)$ by complex conjugation is
the 4-sphere, Math. Ann. 208 (1974) 175--177.

\bibitem{SethiSternZaslow} 
S.~Sethi, M.~Stern and E.~Zaslow,
Monopole and dyon bound states in $N=2$ supersymmetric Yang-Mills 
theories, Nucl. Phys. B457 (1995) 484--510. 

\end{thebibliography}
\end{document}